\documentclass[lettersize,journal]{IEEEtran}
\usepackage{amssymb}
\usepackage{amsfonts}
\usepackage{amsmath}
\usepackage{textcomp}
\usepackage{xcolor}
\usepackage[hidelinks]{hyperref}
\usepackage{graphicx}
\graphicspath{./images/}
\usepackage{epsfig}
\usepackage{epstopdf}
\usepackage{wrapfig}
\usepackage{sidecap}
\usepackage{adjustbox}
\usepackage{amssymb}
\usepackage{color}
\usepackage{array}
\usepackage[caption=false,font=small,labelfont=bf]{subfig}
\usepackage{algpseudocode}
\usepackage{tabularx}
\usepackage[font=footnotesize, labelfont=bf]{caption} 
\usepackage{subcaption}
\usepackage{multirow}
\usepackage{bm}
\usepackage{lipsum}
\usepackage{setspace}
\usepackage{placeins}
\usepackage{afterpage}
\usepackage{etoolbox}
\usepackage{morefloats}
\usepackage{float}
\usepackage{pbox}
\usepackage{booktabs, multirow}
\usepackage{hhline}
\usepackage{varioref}
\usepackage{afterpage}
\usepackage{placeins}
\usepackage{framed}
\usepackage{bibentry}
\usepackage{rotating}
\usepackage[utf8]{inputenc}
\usepackage{csquotes}
\usepackage{algorithm}
\usepackage{algpseudocode}
\usepackage{varwidth}
\usepackage{eqparbox}
\usepackage[symbol]{footmisc}
\usepackage[mathscr]{eucal}
\usepackage{mathbbol}
\usepackage{soul}
\usepackage{mathrsfs}
\usepackage{comment}

\usepackage{enumitem}
\usepackage{dirtytalk}
\usepackage{varwidth}
\usepackage{stfloats}
\usepackage[switch]{lineno}
\usepackage{listings}
\usepackage{array,booktabs,xcolor}

\hypersetup
{
    colorlinks = true,
    linkcolor = blue,
    filecolor = magenta,
    urlcolor = cyan,
    breaklinks = true,
}
\DeclareMathAlphabet{\mathpzc}{OT1}{pzc}{m}{it}

\renewcommand{\algorithmiccomment}[1]{\bgroup\hfill\tiny//~#1\egroup}
\algblockdefx[ForEach]{ForEach}{EndForEach}[1]{\textbf{foreach} #1 \textbf{do}}{\textbf{end foreach}}

\newcommand{\StatexIndent}[1][3]{%
  \setlength\@tempdima{\algorithmicindent}%
  \Statex\hskip\dimexpr#1\@tempdima\relax}
\algdef{S}[IF]{IfNoThen}[1]{\algorithmicif\ #1}%
\makeatletter

\setstcolor{red}

\makeatletter
\newcommand{\multiline}[1]{%
  \begin{tabularx}{\dimexpr\linewidth-\ALG@thistlm}[t]{@{}X@{}}
    #1
  \end{tabularx}
}
\makeatother
\begin{document}\title{Energy-Efficient Routing Protocol in Vehicular Opportunistic Networks: A Dynamic Cluster-based Routing Using Deep Reinforcement Learning}
\author{Meisam Sharifi Sani,~\IEEEmembership{Graduate Student Member,~IEEE,} Saeid Iranmanesh,~\IEEEmembership{Senior Member,~IEEE,}
Raad Raad,~\IEEEmembership{Member,~IEEE,} Faisel Tubbal,~\IEEEmembership{Senior Member,~IEEE}
% <-this % stops a space
\thanks{M.Sharifi Sani, S. Iranmanesh, R. Raad, and F. Tubbal are with the School of Electrical, Computer and Telecommunication Engineering, University of Wollongong, Wollongong, NSW 2522, Australia.
(e-mail: meisam.sharifi@ieee.org; saeidim@uow.edu.au; raad@uow.edu.au; faisel\_tubbal@uow.edu.au)}}%

%\thanks{Manuscript received April 19, 2017; revised August 29, 2017.}}

%\markboth{Journal of \LaTeX\ Class Files,~Vol.~14, No.~2, May~2017}%
%{Shell \MakeLowercase{\textit{et al.}}: Bare Demo of IEEEtran.cls for IEEE Communications Society Journals}

\maketitle

\begin{abstract}
Opportunistic Networks (OppNets) employ the Store-Carry-Forward (SCF) paradigm to maintain communication during intermittent connectivity. However, routing performance suffers due to dynamic topology changes, unpredictable contact patterns, and resource constraints including limited energy and buffer capacity. These challenges compromise delivery reliability, increase latency, and reduce node longevity in highly dynamic environments. This paper proposes Cluster-based Routing using Deep Reinforcement Learning (CR-DRL), an adaptive routing approach that integrates an Actor-Critic learning framework with a heuristic function. CR-DRL enables real-time optimal relay selection and dynamic cluster overlap adjustment to maintain connectivity while minimizing redundant transmissions and enhancing routing efficiency. Simulation results demonstrate significant improvements over state-of-the-art baselines. CR-DRL extends node lifetimes by up to 21\%, overall energy use is reduced by 17\%, and nodes remain active for 15\% longer. Communication performance also improves, with up to 10\% higher delivery ratio, 28.5\% lower delay, 7\% higher throughput, and data requiring 30\% fewer transmission steps across the network.
\end{abstract}


\maketitle

\section{Introduction}\label{Sec_Intro}
\begin{figure}[!t]
    \centering
\includegraphics[width=2.5in]{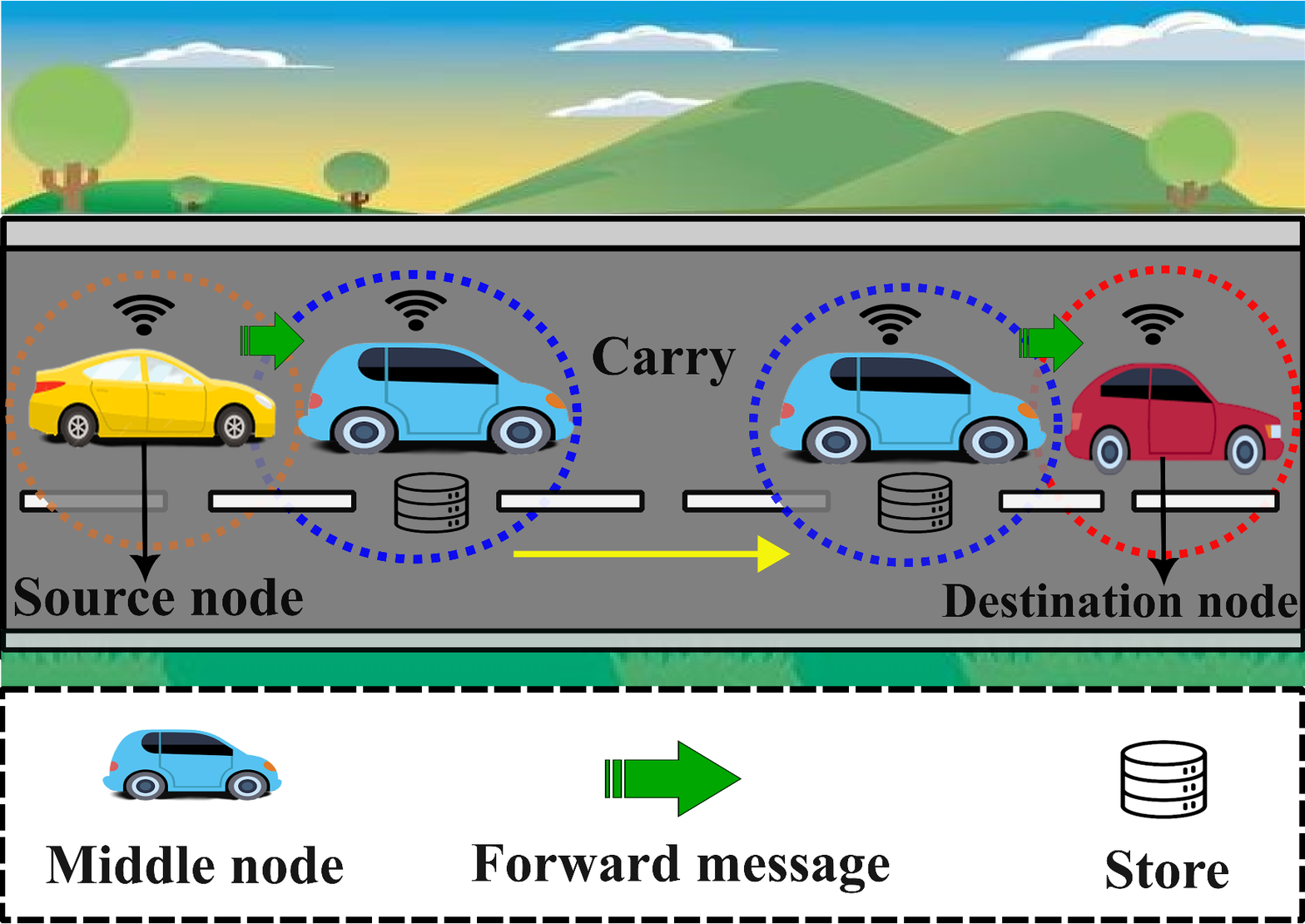}
    \caption{Store-carry-forward architecture in OppNets.}
    \label{Fig_1:Story and forward}
\end{figure}

\IEEEPARstart{S}{mart city} is a concept that aims to integrate information communication with the Internet of Things (IoT) to manage city resources efficiently \cite{A1}. The integration of intelligent vehicles into smart cities can improve safety, traffic flow, and environmental sustainability \cite{A2}. The development of wireless technologies has enabled the implementation of communication systems that rely on vehicles, leading to the emergence of the Vehicular Ad-hoc Network (VANET). VANETs are a type of Mobile Ad Hoc Networks (MANETs) that are differentiated by their highly dynamic network connectivity and frequent topology changes \cite{A3}, \cite{A4}.\\ 
By enabling Intelligent Transport Systems (ITS) to enhance their safety performance and optimize traffic conditions, VANET has established a significant advancement in the field \cite{A5}. This network often leads to a highly dispersed network structure due to the high nodes' mobility that results in intermittent connectivity \cite{A6}. These networks are known as Opportunistic Networks (OppNets) or Delay Tolerant Networks (DTNs). Note that in the paper, we use the term OppNets rather than DTNs for consistency. 
A fundamental communication paradigm in OppNets is the Store-Carry-Forward (SCF) architecture (Figure \ref{Fig_1:Story and forward}). In this model, when a direct communication path between the source and destination is unavailable, data packets are stored locally on a node, carried as the node moves, and then forwarded when the node encounters another that can take the packet closer to its destination. This mechanism either floods the data packets or limit the number of replicas across the network to achieve a high network performance. In the case of flooding, it imposes a high overhead to the network. Moreover, if the buffer of nodes is limited, it increases the packet losses and negatively affecting the delivery ratio. On the other hand, limiting the number of replicas reduces the delivery ratio and increases delays, as data packets have fewer forwarding opportunities.\\ Cluster-based techniques also address these issues by limiting forwarding opportunities to nodes that (i) have a higher capacity to connect with multiple nodes simultaneously, and (ii) possess sufficient buffer space and energy to carry and forward data packets with lower overhead and delays. However, the existing cluster-based methods often struggle with dynamic environments. They lack mechanisms to adapt to rapidly changing network conditions that leads to inefficient cluster head selection, high overhead, and poor energy management \cite{A9}, \cite{A10}, \cite{A11}.

To address the challenges outlined above, it is essential to develop advanced cluster-based routing protocols capable of adapting to dynamic network conditions while minimizing End-to-End delay (E2E delay) path lengths between source and destination nodes. Building on the gaps identified in the literature (Section \ref{Related work}), in this paper, we propose a Cluster-based Routing protocol using Deep Reinforcement Learning called CR-DRL that route the network traffic in a highly dynamic network. Actor-Critic (AC) is the deep reinforcement learning algorithm used in this work due to the ability of learning optimal policies in complex, ever-changing conditions. This adaptive clustering approach dynamically selects cluster heads based on three key real-time metrics: remaining energy, encounter rate, and buffer size. The AC algorithm evaluates these candidates considering performance indicators such as delivery ratio, throughput, and delay to ensure that only the most efficient nodes are chosen as cluster heads. A heuristic function is then employed to refine cluster formation by dynamically adjusting the degree of overlap between clusters based on real-time node density and spatial distribution. This approach ensures that clusters remain optimally structured, balancing connectivity and independence to facilitate efficient data transmission. By strategically linking dispersed nodes, the method transforms an otherwise fragmented network into a well-organized, interconnected structure, making it more compatible with TCP/IP communication protocols. To our knowledge, CR-DRL is the first protocol in the literature that utilizes a deep learning model for cluster-based routing in OppNets. Our contributions and innovations are as follows:
\begin{itemize}
\item We propose a novel routing protocol that incorporates an AC algorithm to intelligently and optimally select cluster heads in OppNets, based on real-time parameters such as encounter rate, residual energy, and buffer size. Unlike existing clustering techniques \cite{A19}, \cite{S1}, \cite{S2}, which often suffer from workload imbalance due to repeatedly selecting the same nodes as cluster heads, our approach dynamically evaluates node capabilities to ensure balanced resource utilization, improved routing stability, and enhanced network performance in highly dynamic vehicular environments. 
\item To address the limitations of fixed-distance clustering in existing methods \cite{New1}, \cite{A19}, \cite{RN1}, \cite{New3}, we propose a heuristic function to dynamically adjust the degree of overlap between clusters based on node density and spatial distribution. This will result in improving both scalability and connectivity.
\item We propose a middleware architecture that enables TCP/IP compatibility in OppNets by transforming dynamic clusters into virtual network segments, using cluster heads as gateways and shared nodes as relays to create stable communication paths without requiring physical infrastructure such as Roadside Units (RSUs).
\end{itemize}

The rest of this paper is organized as follows; Section \ref{Related work} provides a comprehensive literature review to critically analyse the existing gaps. Section \ref{System model} includes the system description and hypotheses. The proposed CR-DRL method and algorithm are presented in Section \ref{Proposed method}. Sections \ref{simulation study} presents parameter simulations and experimental results. Finally, the conclusion and future work are discussed in Section \ref{conclusion and future}.
\FloatBarrier
\section{Related work}\label{Related work}
This section reviews routing strategies in OppNets, focusing on two major approaches: learning-based methods (Reinforcement Learning (RL) / Deep RL (DRL)) and clustering-based methods, highlighting their strengths and limitations.
\vspace{-6pt}
\subsection{Clustering and Machine Learning-Based Routing in OppNets}
In \cite{S1}, the authors propose a RL-based Mobility Adaptive Routing protocol for VANETs that optimizes paths using real-time mobility and traffic parameters, achieving improved delivery performance despite implementation complexity challenges. Similarly, in \cite{S2}, Kumar et al. introduced RLProph, a RL–based dynamic Programming protocol for opportunistic IoT routing that optimizes delivery probability and latency through adaptive decisions on node connectivity, though its applicability is limited by computational overhead. In \cite{S3}, a Latency-Aware Reinforced Routing protocol for OppNets leverages past transmission experiences to adaptively select forwarding paths, thereby reducing E2E delay and improving delivery in dynamic topologies; however, its continuous learning and frequent state updates impose significant computational and energy overhead, limiting suitability for resource-constrained nodes. 

In \cite{A19}, the authors introduce QUERA, which integrates Q-learning into the Routing Protocol for Low-Power and Lossy Networks (RPL) to enable adaptive parent selection based on energy, link quality, and mobility. This design enhances packet delivery and reduces delay, particularly in mobile IoT environments. The method requires Q-table maintenance, which adds some communication and computation overhead, yet this trade-off highlights an interesting balance between adaptability and resource efficiency that could inspire further optimization. A Reliable and Mobility-Aware RPL (ARMOR) is proposed in \cite{New2} that incorporates a Time-to-Reside (TTR) metric derived from velocity and positional data to better estimate connectivity duration. This mechanism significantly improves packet delivery in low-mobility scenarios. Its evaluation is mainly focused on such environments, leaving opportunities for future exploration of its applicability in higher-mobility conditions, such as vehicular networks, and for strategies to manage the added control signalling more efficiently.

In \cite{New1}, the authors propose Energy Efficient Hybrid Clustering and Hierarchical Routing (EEHCHR), which employs Fuzzy C-Means (FCM) clustering based on residual energy and node-to-base station distance to form clusters with both direct and central cluster heads. This design effectively reduces overhead and extends network lifetime. The approach, however, is primarily evaluated under static node deployment, which may be less representative of highly dynamic OppNets scenarios, offering an opportunity for further exploration. In another work \cite{N4}, the Energy-efficient technique for selection of Optimum Number of Cluster Head and Grid Head (EOCGS) protocol proposes a hybrid clustering approach that also leverages FCM, residual energy, and base station distance to selectively form clusters, thereby reducing overhead. While effective, the protocol’s reliance on predefined grid structures means its adaptability to changing or mobile environments could be further investigated. A modified version of Residual Energy based Cluster Head Selection (MRCH) in Low Energy Adaptive Clustering Hierarchy (RCH-LEACH) is proposed in \cite{N5}, which improves energy efficiency by selecting cluster heads based on node energy levels and optimizing the cluster head percentage. MRCH extends network lifetime and shows robustness under different base station placements. Its cluster-rotating mechanism is well suited for stable deployments, while future work could consider enhancements for scenarios with higher mobility and rapidly evolving topologies.
A Cluster-based Routing using Meta-Heuristic Algorithms for VANET (CRMHA-VANET) is proposed in  \cite{New3}, which combines Gray Wolf Optimization (GWO) for cluster head selection and Technique for Order of Preference by Similarity to Ideal Solution (TOPSIS) for route selection, optimizing intra-cluster distance, link quality, and load balancing. Although it achieves good performance, its dependence on RSUs limits scalability in OppNets with sparse infrastructure. In \cite{RN3}, a cluster-based routing scheme employing Harris Hawks Optimization (HHO) is proposed, where cluster head selection is guided by node mobility, residual energy, and transmission distance. This approach enhances the delivery ratio and reduces latency. HHO-based clustering offers clear benefits for cluster-head selection and load balancing, though factors such as computational overhead and adaptability in dynamic environments remain areas for further consideration. 

Moghaddasi et al. \cite{RN1} propose a Double Deep Q-Network (DDQN) for 5G-enabled Vehicular Edge Computing (VEC), which reduces energy consumption and latency through adaptive task offloading. However, its reliance on stable edge infrastructure suggests opportunities for further evaluation in highly dynamic environments such as OppNets. In \cite{RN2}, a DRL-based task offloading method is proposed for three-layer Device-to-Device (D2D)–edge–cloud IoT architecture, employing advanced neural networks to optimize offloading decisions across heterogeneous environments. The approach achieves significant improvements in energy efficiency, latency reduction, and utility enhancement through adaptive real-time decision-making. However, future research could explore lightweight implementation strategies to optimize computational requirements for broader deployment across diverse IoT infrastructures. In \cite{A41}, the authors propose ECRDP, an Efficient Clustering Routing approach that integrates Density Peaks Clustering (DPC) with Particle Swarm Optimization (PSO) to adapt routing decisions according to vehicle density, traffic conditions, road topology, and link quality. While the method demonstrates adaptability, its high computational complexity results in suboptimal delivery ratios and throughput.

Existing protocols face challenges in cluster head selection, delivery ratio, E2E delay, and energy efficiency, particularly in highly dynamic OppNets. RL-based methods \cite{S1}, \cite{S2}, \cite{S3}, \cite{A19}, and \cite{New2} offer adaptability but incur computational overhead, while clustering approaches \cite{New1}, \cite{New3}, \cite{RN3}, \cite{N4}, \cite{N5}, \cite{A41}, and \cite{RN1} struggle with static thresholds or infrastructure dependency. CR-DRL addresses these gaps by integrating an AC framework for adaptive cluster head selection and a heuristic function for dynamic cluster overlap, enhancing scalability, energy efficiency, and compatibility with TCP/IP protocols without heavy infrastructure reliance. Table \ref{RelatedWork} provides a detailed comparison of these methods’ strengths, limitations, and metrics.

\begin{table*}[!t]
    \centering
    \caption{Comparison of the various clustering and machine learning routing protocols}\label{RelatedWork}
      \begin{tabular}{|p{2.5cm}|p{5.2cm}|p{5.2cm}|p{4cm}|}
    \hline
    \textbf{Method (Ref.)} & \textbf{Strengths} & \textbf{Limitations} & \textbf{Performance Metrics} \\
   \hline
\cite{A19} Q-learning &Adapts to mobility; improves energy efficiency, packet delivery, and reduced delay.& Requires Q-table maintenance; increased memory and processing demand.&  Packet delivery ratio, delay.\\
\hline
\cite{S1} RL&Dynamic path learning; better delivery in VANETs. & High training overhead; complex reward design.& Delivery ratio, latency. \\
\hline
\cite{S2} RL&Effective in intermittent links; improves delivery. & Computationally intensive; higher overhead.&  Delivery rate, latency.\\
\hline
\cite{New1} FCM &Balanced energy use; lower overhead; extended lifetime. & Static thresholds; lacks adaptability.& Network lifetime, energy consumption.  \\
\hline
\cite{RN1} DDQN &Adaptive offloading; reduced energy and latency. & Assumes stable edge infrastructure; unsuitable for OppNets.&Energy consumption, latency. \\
\hline
\cite{New3} GWO& Multi-objective fitness function; hierarchical structure for stability; balanced clustering.&Increased packet overhead; RSU dependency; scalability concerns.&Delivery rate; throughput; E2E delay. \\
\hline
\cite{RN3} HHO&Improved delivery ratio; reduced delay. & Unstable convergence; needs parameter tuning.&Delivery ratio, delay.\\
\hline
\cite{N4} FCM& Balanced energy use; prolonged network life. & Grid structure limits adaptability.& Energy usage, network lifetime\\
\hline
\cite{N5} Clustering & Lifetime extension; robust under various BS locations. & Static rotation; not suited for mobile/dynamic cases.& Energy efficiency, network lifetime.\\
\hline
\cite{A41} DPC + PSO& Context-aware routing; handles density/topology variations. & High complexity; underperforms in delivery ratio.& Delivery ratio, throughput.\\
\hline
\cite{S3} RL & Reduces E2E delay; adapts to topology changes. & High learning cost; not ideal for constrained devices.& E2E delay, delivery rate.\\
\hline
\cite{New2} TTR & Improved delivery ratio; stable connection establishment; comparable power consumption.& Control overhead increase; limited to low-speed mobility; potential latency increase.&Delivery rate; power consumption, E2E delay. \\
\hline
\cite{RN2} DRL &Improved energy efficiency; latency reduction and utility enhancement.& Complexity; high learning cost&Delivery ratio, delay, energy efficiency.\\ 
\hline
    \end{tabular}
\end{table*}
\section{System model}\label{System model}
\vspace{-6pt}
\subsection{Preliminary}\label{Preliminary}
In this section, we present the RL strategy adopted for routing in OppNets, focusing on the AC framework \cite{AC1}, an advanced variant of the policy gradient with a baseline algorithm. This framework effectively tackles the challenges of dynamic environments like OppNets, where traditional routing techniques often fail due to high node mobility and intermittent connectivity. The AC framework is formulated as a Markov Decision Process (MDP), represented by the tuple ($s,a, P,R,\gamma$) \cite{N1}, where:
\begin{itemize}
    \item $s$ represents the state space of the environment.
    \item $a$ is the action space defining the possible decisions of the agent.
    \item $P$ defines the state transition probability, influenced by node mobility and network dynamics.
    \item $R$ represents the reward function, providing feedback for each action taken.
    \item $\gamma$ is the discount factor, determining how much future rewards influence the current decision ($0\leq\gamma\leq1$).
\end{itemize}
The global state (s) is constructed from the aggregation of individual node states, $s=\{{s_1,s_2,...,s_N}\}$ where $s_i$ represents the state of the network at time step $t$ of node $i$. This state provides the necessary context for both the actor, which selects actions, and the critic, which evaluates the quality of those actions. The objective function in RL aims to maximize the expected cumulative discounted reward: \begin{equation}\label{Eq2_GeneralF} J(\theta)= \mathbb{E}_{{{\pi}_\theta}_{(a|s)}} \left[ \sum_{t=0}^{\infty} \gamma^t R_t \right] \end{equation} 
\noindent Where $J(\theta)$ is the performance objective, ${\pi}_\theta(a|s)$ represents the policy parameterized by $\theta$, which defines the probability of taking action $a$ in state $s$ and $R_t$ is the reward received at time step $t$. \cite{A50}. The policy gradient theorem \cite{N2} provides the gradient of the objective function:
\begin{equation}\label{Eq4_PolicyG} \nabla_\theta J(\theta)= \mathbb{E}{{\pi}_\theta} \left[\nabla_\theta \log {\pi}_\theta(a|s) Q^\pi (s,a) \right] \end{equation}
\noindent Where $Q^\pi (s,a)$ is the action-value function, representing the expected return after taking action $a$ in state $s$ under policy $\pi$. While the above gradient works, it often suffers from high variance, especially in dynamic environments such as OppNets. To mitigate this issue, a baseline function $b(s)$, typically the state-value function $V(s;\omega)$, is proposed \cite{A51}:
\begin{equation}\label{Eq5_SataeV} \nabla_\theta J(\theta)= \mathbb{E}{{\pi}_\theta} \left[\nabla_\theta \log {\pi}_\theta(a|s) \left(Q^\pi (s,a)-V(s;\omega)\right) \right] \end{equation}
\noindent Where $Q^\pi (s,a)-V(s;\omega)$ is the advantage function, measuring how much better an action is compared to the baseline \cite{A51}:
\begin{equation}\label{Eq6_Baseline} A(s,a)= {Q}^\pi (s,a)-V(s;\omega) \end{equation}
\noindent The actor updates the policy parameters $\theta$ using:
\begin{equation}\label{Eq7_ActorU} \theta\leftarrow\theta+{\alpha_{a}} A(s,a) \nabla_\theta \log {\pi}_\theta(a|s) \end{equation}
\noindent Where $\alpha_{a}$ is the actor’s learning rate, and $A(s,a)$ guides the update based on the advantage of the chosen action. The critic updates the value function parameters $\omega$ using the Temporal-Difference (TD) error $\delta_{t}$ \cite{A38}:
 \begin{equation}\label{Eq8_TD} \delta_{t}=r+\gamma V(s';\omega)-V(s,\omega) \end{equation}
 \noindent Where $r$ is the reward after taking action $a$ in state $s$ and transitioning to $s'$. The critic’s update rule is:
 \begin{equation}\label{Eq9_CriticU} \omega\leftarrow\omega+\alpha_c \delta_t \nabla_\omega V(s;\omega) \end{equation}
 \noindent Where $\alpha_c$ is the critic’s learning rate.\\
 \begin{comment}
      While advanced deep reinforcement learning (DRL) methods such as Trust Region Policy Optimization (TRPO) \cite{P1} and Proximal Policy Optimization (PPO) \cite{P2} improve training stability, they cause additional computational complexity and require extensive hyperparameter tuning. This makes them impractical in resource-constrained environments such as OppNets. In contrast, the AC framework offers a more computationally efficient solution, and balancing:
 \begin{itemize}
     \item Policy-based methods (Actor): Enables rapid adaptation to changing network conditions.
     \item Value-based methods (Critic): Provides feedback, reducing variance and improving stability.
 \end{itemize}
This dual mechanism allows the AC framework to efficiently balance exploration (trying new strategies) and exploitation (leveraging successful strategies), ensuring adaptability in dynamic OppNets topologies.
\end{comment}
\vspace{-0.2in}
\subsection{System description}
Consider a city in which ${n}$ vehicles traverse from one location to another in accordance with a mobility model based on maps. In this context $v=\{{v}_{1},{v}_{2},{v}_{3},...,{v}_{n}\}$ denotes a set of vehicles, each equipped with a Wi-Fi module enabling them to initiate, buffer, transmit, and receive messages, and ${v}_{i}$ represents ${i}$th vehicle in the network. The communication network among these vehicles is established wirelessly, devoid of any physical infrastructure.\\ Initially, all vehicles possess the same transmission range ($\beta_{v_i}\leq\rho$). $ \rho $ represents the maximum transmission range, which varies across different scenarios. Consideration is given to the fact that two nodes (${v}_{i},{v}_{j}$) can only communicate when they are within their communication ranges. In other words, we consider the transmission range of ${v}_{1}$ and ${v}_{2}$ as $\beta_{{v}_{1}}$ and $\beta_{{v}_{2}}$, respectively, and the shared area as $\beta_{{v}_{1}}\cap \beta_{{v}_{2}}\neq 0$. Furthermore, in cases where the node ${v}_{i}$ is unable to communicate with the other nodes ($\beta_{{v}_{i}}\cap \beta_{{v}_{j}}=0$) it resorts to the SCF architecture shown in Figure \ref{Fig_1:Story and forward} until it meets the other nodes. Every node (${v}_{i}$) can replicate and transmit a message upon encountering another node when $\beta_{{v}_{i}}\cap \beta_{{v}_{j}}\neq 0$. This means that before clustering, in every interaction with other nodes, a node is able to send a message without any restrictions. Each node sends out a beacon message at fixed time intervals (every 1 second). Neighboring nodes that receive these beacons update their knowledge about the local network topology. \\ We assume that every node ($v_i$) is equipped with a GPS receiver, enabling universal access to GPS data for all nodes, simplifying distance calculations between them. This assumption is based on each vehicle's ability to independently receive GPS signals directly from satellites, rather than depending on the transmission of GPS data between nodes. The GPS provides the geographical coordinates of the node, and the coordinates can be treated in a $2D$ space as (${x_i},{y_i}$).\\ The city is divided into $k$ number of clusters as denoted by $C=\{{C}_{1},{C}_{2},{C}_{3},...,{C}_{K}\}$, and $C_j$ represents the $j$th cluster in network. Each cluster ${C}_{j}$ has a cluster head that is shown by ${\chi}=\{{\chi}_{1},{\chi}_{2},{\chi}_{3},...,{\chi}_{i}\}$, where ${\chi}_{i}$ represents ${i}$th cluster head.  To put it differently, every cluster (${C}_{j}$) has a cluster head (${\chi}_{i}$) that is a member of the (${\chi}_{i}\in {C}_{j},\forall_{j}\in [1,K]$) same cluster. All nodes (${v}_{i}$) in each cluster (${C}_{j}$) are allowed to send and receive information with their cluster head (${\chi}_{i}$) only ($\forall{v}_{i}\in {C}_{j},{v}_{i}\longleftrightarrow {\chi}_{i}$). Each cluster has a transmission range ($\beta_{C_j}$) which is the same as the transmission range of the cluster head ($\beta_{\chi_i}=\beta_{C_j}$), where nodes can easily communicate with their cluster head. Moreover, if the node (${v}_{i}$) is unable to establish communication with the cluster head ($\beta_{{\chi_{i}}}\cap \beta_{{v}_{i}}=0$), it will extend its transmission range until it connects with its cluster head. \\ Every 100 ms, each cluster head (${\chi}_{i}$) broadcasts a unique beacon frame to all nodes (${v}_{i}$) within its coverage area. This enables each node within a cluster to recognize its respective cluster head. Once a new node joins the cluster, it can easily identify the cluster head and transmit information to it. The cluster head is responsible for collecting local data from its member vehicle nodes, aggregating them, and sending them to other common members. Also, the cluster head keeps a record of its cluster members in a table that is regularly updated whenever a new member is discovered. We consider the node that has the probability of interacting with other nodes and the most remaining energy and the maximum buffer size as the candidate cluster head, because it indicates that the node is located in an area with high density of nodes and has enough energy to communicate with other nodes, and also, has enough buffer for receive packet from other nodes, this will be discussed in details in section \ref{Proposed method}.\\ Let us assume $\aleph=\{{\aleph}_{1},{\aleph}_{2},{\aleph}_{3},...,{\aleph}_{q}\}$ represent a group of common members that are communicated between clusters ($\forall{\chi}_{i}\in {C}_{j}, \forall{q}\in{\aleph}, {v}_{i}\in {C}_{j},{v}_{i}\longleftrightarrow{\chi}_{i},{\chi}_{i}\rightarrow{\aleph}_{q}$), and ${\aleph}_{q}$ represents ${q}$th common member in the network. The ${\aleph}_{q}$ is responsible for transmitting packets from ${C}_{i}$ one cluster to ${C}_{j}$ another, and also, cluster head ($\chi_{i}$) can not be a common member ($\aleph_{q}$) with other clusters ($\chi_{i}\neq\aleph_{q}$). 
\begin{table}[!ht]
    \centering
   \caption{Notation used in the CR-DRL method}\label{tab_1}
       \begin{tabular}{|p{1.8cm}|p{6.3cm}|}
    \hline
 \textbf{Notation}& \textbf{Describe} \\
\hline
$s$, $a$ & state value, and action value. \\ 
\hline
$R$, $\gamma$, $\alpha_a$, $\alpha_C$& Reward, Discount factor, Learning rate for Actor, and Learning rate for Critic\\ 
\hline
$\theta$, $\omega$, $\delta_t$, $\varepsilon$  &Weight, The parameters (weights) of the neural network, Temporal-Difference (TD) error, and Entropy term.\\
\hline
$V(s;\omega)$, $Q^\pi(s,a)$& state-value function and action value function.\\
 \hline
$v$, $\beta_{v_i}$&Vehicle, and Vehicle's transmission range\\
\hline
$C$, $\beta_{C}$, $\rho$&Cluster, Cluster transmission range and Maximum vehicle transmission range\\
\hline
${\chi}$, $\beta_{\chi}$, $\aleph$& Cluster head, Cluster head transmission range, and Common member\\
\hline
$S$, $D$, $P$, $Z$ &Source, Destination, Path, and Route\\
\hline
$G_i$, $\tau_{t}$, ${T}_{e}$& A link between vehicles, The reliability of the link, and Estimated availability time\\
\hline
$H$, $d$, $N$, $U$ & Node density, Density, Number of nodes, and Network area\\
\hline
$\varrho_d$, $\Delta$&ADT function, and Scaling factor\\
\hline
${\phi}_{{v_i}_{\max}}$, ${\phi}_{v_i}$, ${\phi}_{{v_i}_{(n)}}$, ${\phi}_{{v_i}_{(c)}}$ &Maximum encounter rate, Encounter rate, New encounter rate, Current encounter rate, and minimum energy for selecting cluster head. \\
\hline
$CW$, $\lambda$&Current window, and Determined factor\\
\hline
$\Psi_{Initial}$, $\Psi_{{v}_{i}}$, $\Psi_{(t)}$, $\eta_{{v}_{i}}$& Initial energy, Remaining energy,The energy of node in time slot ${t}$, and Maximum buffer size \\
\hline
$B_{max}$, $B_{(t)}$&Maximum number of messages and Number of messages currently stored in the buffer\\
\hline
${T}_{AC}$, ${T}_{Clustering}$, ${T}_{Routing}$ $T_{Total}$ &AC algorithm  complexity, Clustering complexity, Routing complexity, and Total complexity \\
\hline
$M_1$, $M_2$, and $M_3$ & Normalized weight.\\
\hline
    \end{tabular}
\end{table}
In our network, every node utilizes a Multiple Input Multiple Output (MIMO) channel to enhance communication quality among vehicles. When a vehicle is a common member of multiple clusters, it can establish communication with other cluster heads. To ensure higher transmission speed and reliable communication between vehicles, we have implemented MIMO $4$ in this system \cite{N3}. Lastly, clusters should have minimal communication with other clusters to facilitate communication through shared members. In Table \ref{tab_1}, we have provided a summary of the notations used in the CR-DRL method.
\vspace{-0.2in}
\subsection{Problem formulation}
In vehicular networks, data can be transferred from a source (${S}$) to a destination (${D}$) through multiple routing paths. In a high-traffic urban area, the availability of links between vehicles fluctuates due to frequent topology changes, necessitating an adaptive routing strategy. To model this process, consider a route ${Z}$ comprising ${n}$ communication links that connect intermediate nodes between ${S}$ and ${D}$.\\
\begin{equation}\label{Eq_10}
\centering
    {G}_{i}={G}_{1}=({S},{v}_{1}), {G}_{2}=({v}_{1},{v}_{2}),....,{G}_{n}=({v}_{n},{D})
\end{equation}
We suppose ${G_{i}}$ is a link between vehicle ${V}_{i}$ and ${V}_{j}$ at time ${t}$. All intermediate links ${G}_{i}$ have particular reliability, which corresponds to the following:
 \begin{equation}\label{Eq_11}
     \centering
     \tau_{t}(G_{i})=\int_{t}^{t+{T}_{e}}f(T)dt ~if~ {T}_{e}>0
 \end{equation}
 $\tau_{t}$ is the reliability of the link ${G_{i}}$ at the time ${t}$. ${T}_{e}$ is estimated availability time, and ${f}(T)$ is the probability density function of communication \cite{A28} duration ${T}$ of ${v}_{i}$ and ${v}_{j}$. Traditional routing strategies struggle to ensure reliable data transmission in such an environment, making it crucial to develop an adaptive approach that maximizes delivery efficiency while minimizing energy consumption and transmission delays. To address this challenge, we frame the routing problem as an optimization task that seeks to enhance network performance while maintaining resource efficiency. To optimize data transmission in vehicular networks, we maximize delivery ratio and throughput, which ensure efficient data forwarding, while minimizing E2E delay to reduce transmission latency. The optimization problem is formulated as follows:
\begin{equation}
    \centering
    \max (F(Z))=  (\hat{\mu}) + (\hat{\digamma}) - (\hat{\mathbb{D}})
\end{equation}
Subject to:
  \begin{equation}\label{linkA}
\tau(Z(S,D)) = \prod_{i=1}^{n} \tau_t(G_i) \ where \ G_i \in Z(S,D)
\end{equation}
\begin{equation}\label{EnergyP}
\Psi_{{v}_{i}} \geq E_{threshold}, \quad \forall v_i \in Z(S,D)
\end{equation}
\begin{equation}\label{BufferP}
 {\eta}_{v_i} \leq B_{max}, \quad \forall v_i \in Z(S,D)
\end{equation}
Eq. \ref{linkA} represents the cumulative reliability of all links in the route, where each $\tau_t(G_i)$ corresponds to the reliability of an individual link $G_i$. The product operation ensures that the overall route reliability decreases proportionally with weaker links. In Eq. \ref{EnergyP} $(\Psi_{{v}_{i}})$ represents the residual energy of a node at time $t$, which determines its ability to participate in data forwarding and network operations. Since vehicles have finite battery reserves, nodes with insufficient energy may disrupt transmissions or drop out of the network. By enforcing $E_{threshold}$, we ensure that only nodes with sufficient energy participate in forwarding, thereby prolonging network lifespan.\\ In Eq. \ref{BufferP} $\eta_{{v}_{i}}$ represents the maximum buffer size, if buffers exceed $B_{max}$, packets may be dropped, leading to data loss. By enforcing this constraint, we ensure that nodes manage their storage efficiently, preventing excessive queuing delays and avoiding packet drops, which would otherwise degrade network performance.\\ While these constraints define an optimal routing framework, traditional approaches struggle to adapt in real time. To address this, we propose CR-DRL as a reinforcement learning-based model that continuously learns optimal routing strategies, dynamically selects cluster heads, and adapts to evolving network conditions.
\section{Proposed method}\label{Proposed method}
Clustering in OppNets organizes vehicles into groups based on parameters such as distance and location to enhance information exchange and communication. Various algorithms optimize this process \cite{A29}, \cite{A30}, \cite{A31}, \cite{A32}, but selecting an effective cluster head either centrally via infrastructure or distributively among vehicles remains a key challenge \cite{A33}. In dynamic vehicular networks, traditional clustering struggles with rapid topology changes, causing packet loss and routing inefficiencies. RL offers a solution by enabling adaptive, real-time decision-making. To address these challenges, we propose CR-DRL, an AC based clustering and routing approach designed to optimize communication efficiency in OppNets.
CR-DRL approach leverages the strengths of the AC algorithm to enhance the efficiency and effectiveness of cluster head selection. Initially, we compute the Manhattan distance between each node and its neighbors using GPS data. We assume that $d$ denotes the Manhattan distance between node $v_i$ and node $v_j$ (${d}={|{x}_{i}-{x}_{j}|}+{|{y}_{i}-{y}_{j}|}$). Subsequently, through the definition of three states involving the highest encounter rates (${\phi}_{{v_i}_{\max}}$), remaining energy levels ($\Psi_{{v}_{i}}$), and buffer size ($\eta_{{v}_{i}}$), the selection of candidate cluster heads will be determined. \\ We utilized the maximum encounter value of neighboring nodes within a specified range. To monitor a node’s encounter frequency, CR-DRL tracks two local variables: encounter history (${\phi}_{v_i}$) and a current window counter ($CW$). The encounter history (${\phi}_{v_i}$) represents the node's historical encounter rate, calculated using a weighted moving average. Meanwhile, the $CW$ records the number of encounters within the current time interval. Periodically, the encounter history (${\phi}_{{v_i}_{(n)}}$) is updated to include the latest $CW$ values, ensuring that the current encounter rate (${\phi}_{{v_i}_{(c)}}$) is accurately reflected. The update calculations for the encounter history (${\phi}_{v_i}$) are performed as follows:
\begin{equation}\label{Eq_14:ER}
    \centering
    {{{\phi}_{{v}{i}}}_{(n)}}=\lambda*CW+(1-\lambda)*{{{\phi}_{{v}_{i}}}_{(c)}}
\end{equation}
The exponentially weighted moving average prioritizes the most recent complete ${CW}$, with the level of emphasis determined by the factor $\lambda$. Updating the ${CW}$ is straightforward: it increments with each encounter. At the end of each update interval, the encounter history is recalculated, and the ${CW}$ is reset to zero. Our experiments showed that a $\lambda$ value of $0.85$ works well. We found that if the interval time is too long, the encounter rate becomes an unbalanced value, whereas if the interval time is too short, the node does not have enough opportunities to encounter neighboring nodes. In our tests, an update interval of around 30 seconds provided the best balance.\\ The goal of this optimization problem is to maximize the total weights of the links between a node (${v}_{i}$) and its neighbors. This means that each vehicle keeps a list of neighboring nodes, and it is preferable to select a cluster head with the highest ${\phi}_{v_i}$ for effective communication. The highest ${\phi}_{v_i}$ becomes a key factor in choosing a cluster head. We denote ${\phi}_{{v_i}_{max}}$ as the encounter history of node ${v_i}$, representing the number of encounters node ${v_i}$ has had with its neighbors, which can be expressed as follows:
\begin{equation}\label{Eq_17:maxER}
    \centering
           {\phi_{{v}_{i}}}_{\max}={arg\max_{{v}
           _{i}\in {v}}({\phi_{{v}_{i}}})} 
    \end{equation}
The residual energy denoted as ($\Psi_{{v}_{i}}$) is another factor that delineates the amount of energy accessible within a node's battery. This factor, which will henceforth be referred to as the residual energy of ${v}_{i}$, can be defined as follows: 
\begin{equation}\label{Eq_18:maxRE}
    \centering
\Psi_{{v}_{i}}=\frac{\Psi_{(t)}}{\Psi_{Initial}}
\end{equation}
Where $\Psi_{(t)}$ denotes the energy of node ${v}_{i}$ in time slot ${t}$. In addition, in the CR-DRL method, a node is eligible for cluster head selection only if its residual energy is at least 25\% of its initial energy ($E_{threshold} \geq 0.25*\Psi_{Initial}$). This threshold ensures that selected cluster heads have sufficient energy to handle data forwarding, intra-cluster coordination, and inter-cluster communication without rapidly depleting their resources. After conducting extensive simulation studies, we observed that setting $E_{threshold}$ to 10\% of the initial energy leads to frequent cluster head changes due to rapid energy depletion among selected nodes. This instability results in increased network overhead, higher re-clustering frequency, and reduced overall efficiency. In contrast, when $E_{threshold}$ is set to 50\%, a significant portion of nodes with sufficient energy remains underutilized, leading to inefficient resource allocation. This restriction reduces the number of eligible cluster heads, potentially causing longer routing paths and increased transmission delays, ultimately compromising network scalability and adaptability in dynamic environment. \\
Last factor is maximum buffer size ($\eta_{{v}_{i}}$). The maximum buffer size typically refers to the maximum number of messages that a node's buffer can hold at any given time. A cluster head with a larger buffer space is a better candidate for message delivery, so one of the important factors for choosing a cluster head can be the maximum buffer size ($\eta_{{v}_{i}}$). $\eta_{{v}_{i}}$ according to the following definition:
\begin{equation}\label{Eq_19:BS}
    \centering
    {\eta}_{v_i}= \frac{B_{(t)}}{B_{max}}
\end{equation}
Where $B_{max}$ is maximum number of messages the buffer can hold, and $B_{(t)}$ is number of messages currently stored in the buffer at time (${t}$).\\ 
In CR-DRL method, the state (${s}$) in AC algorithm consists of information about the nodes, including their maximum encounter rate, remaining energy, and buffer size. ${s}_{i}=(\phi_{1}, \phi_{2},...\phi_{n},\Psi_{1},\Psi_{2},...\Psi_{n},\eta_{1},\eta_{2},...\eta_{n})$. Where $n$ is the number of nodes in the network.
In the CR-DRL method, the reward ($R$) is designed by normalizing three factors to balance the delivery ratio ($\hat{\mu}$), throughput ($\hat{\digamma}$), and E2E delay ($\hat{\mathbb{D}}$), while also addressing scaling issues:
\begin{equation}\label{Eq3_Reward}
    \centering
    R=M_1.({\hat{\mu}})+M_2.(\hat{\digamma})-M_3.(\hat{\mathbb{D}})
\end{equation}
Where $M_1$, $M_2$, $M_3$ are scaling coefficients ensuring balanced contributions. Then, metrics are normalized to prevent dominance of one metric over others. We assign equal importance to each of these metrics ($M_1, M_2, M_3=1$). By giving equal weights to these three metrics, we ensure a balanced approach that does not overly prioritize one aspect at the expense of others. 

\begin{figure}[!t]
    \centering
\includegraphics[width=2.5in]{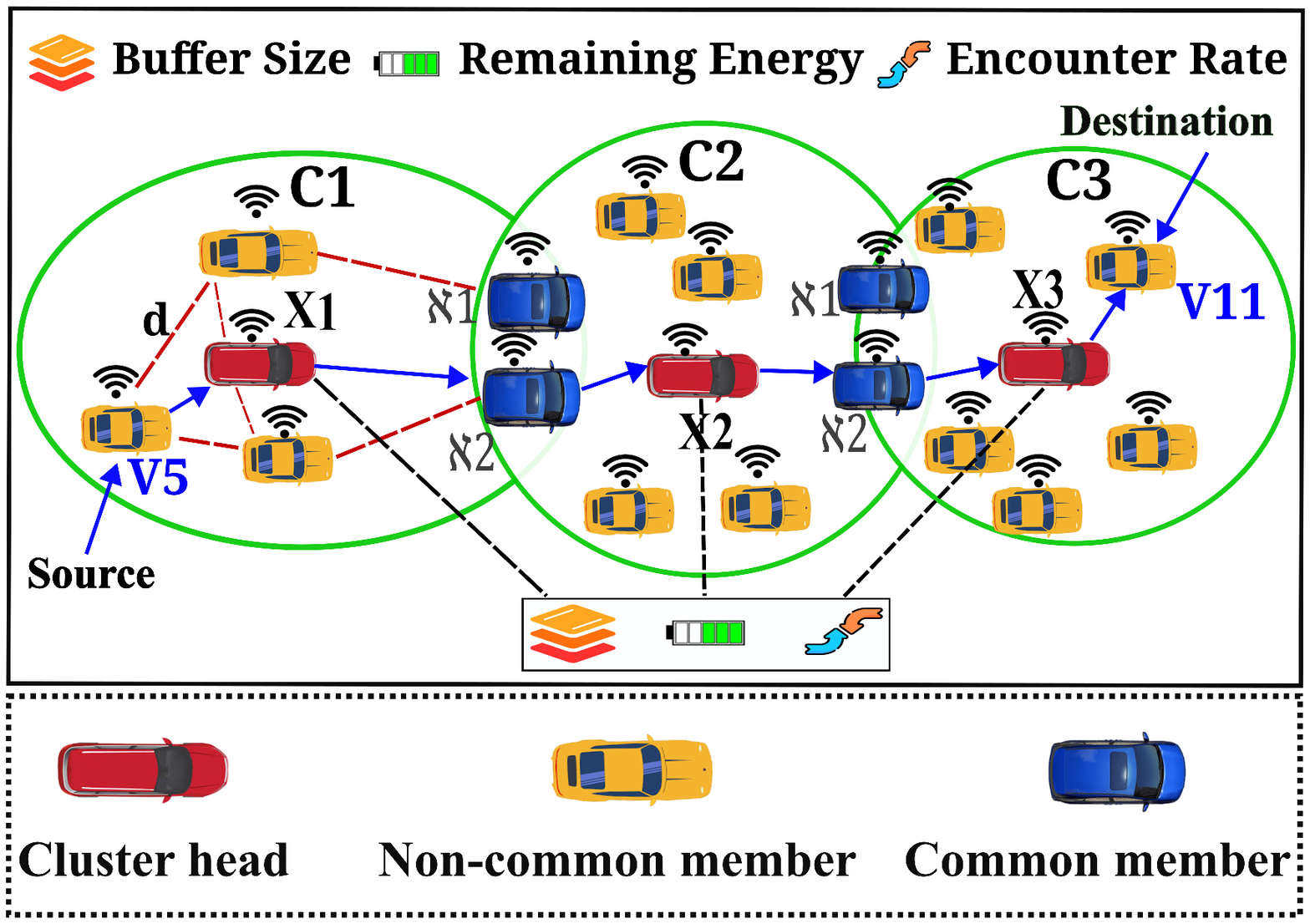}
    \caption{Dynamically clustered routing using common members.}
    \label{fig_2:Routing}
\end{figure}

In addition, the potential instability and convergence issues commonly associated with DRL methods have been mitigated through key strategies. First, we employ the AC algorithm with entropy regularization, which helps maintain a balance between exploration and exploitation during training. This reduces the likelihood of the policy getting stuck in suboptimal states and promotes stable learning dynamics.The policy loss function with entropy regularization is defined as:
\begin{equation}\label{ANFunction}
    \centering
    L(\theta) = -\mathbb{E}_{s,a \sim \pi_{\theta}} \left[ \log \pi_{\theta}(a|s) A(s,a) \right] +  \varepsilon O(\pi_{\theta})
\end{equation}
Where $O(\pi_{\theta})$ = $\sum \pi_{\theta}(a|s) \log \pi_{\theta}(a|s)$) is the entropy term, and $\varepsilon$ controls the influence of the entropy term. Second, we implement an experience replay buffer, which allows the model to learn from past routing decisions while avoiding overfitting to recent, possibly transient network conditions. By storing and randomly sampling past experiences, the model gains a more generalized understanding of network dynamics, improving adaptability to frequent topology changes. The parameter update rule is:
\begin{equation}
    \centering
    \theta \leftarrow \theta + \alpha \nabla_{\theta} \mathbb{E}_{(s, a, r, s') \sim \mathcal{Y}} \left[ \log \pi_{\theta}(a|s) A(s,a) \right]
\end{equation}
Where ($\mathcal{Y}$) is the replay buffer, and $\alpha$ is the learning rate. Third, adaptive learning rate scheduling is utilized to fine-tune the training process dynamically. By reducing the learning rate when performance plateaus or fluctuates significantly, we improve convergence stability. Lastly, we apply gradient clipping to prevent exploding gradients, which can destabilize the training process. Together, these mechanisms ensure that CR-DRL achieves stable performance from the early stages of training, minimizing packet loss and routing inefficiencies typically observed in DRL-based methods. By integrating entropy-driven exploration, replay-based learning, and adaptive optimization, CR-DRL remains robust even in networks with highly unpredictable node behaviors.\\
\begin{comment}
  Once the cluster heads are chosen, the inevitable uncertainty caused by estimation errors in contact probabilities and the unpredictable patterns of collisions among moving nodes make it likely that multiple clusters of different sizes will form. In high-density areas, some of these clusters may overlap significantly. This means that adjacent clusters will share a substantial number of common members, while in low-density areas, distant clusters will have no members in common ($\tau_{t}(G_{i}=0$). Therefore, creating precise clusters necessitates considering a variety of factors.  
\end{comment}
After selecting cluster heads, uncertainty from estimation errors and unpredictable collisions often leads to clusters of varying sizes, with significant overlap in high-density areas and no overlap in low-density ones. Thus, precise clustering requires accounting for multiple factors. To address this challenge, we propose a heuristic function known as the Adaptive Distance Threshold (ADT), designed to evaluate the distance between two clusters in order to achieve an optimal degree of overlap. The clustering process dynamically adjusts the maximum allowable distance between nodes to form clusters based on the current network conditions, such as node density and distribution. The node density ($H$) is calculated as the ratio of the total number of nodes ($N$) to the area of the network ($U$):
\begin{equation}\label{Eq_13:Density} \centering H = \frac{N}{U} \end{equation}
For a uniformly distributed network, the average distance ($d$) between nodes can be approximated as:
\begin{equation}\label{Eq_14Distance} \centering d \approx \frac{1}{\sqrt{H}} \end{equation}
The ADT, denoted as $\varrho_d$, is calculated as:
\begin{equation}\label{Eq_15:ADT} \centering \varrho_d = \Delta d = \Delta \left(\frac{1}{\sqrt{H}}\right) \end{equation}
Here, $\Delta$ is a scaling factor that adjusts the sensitivity of the threshold to changes in density.
\begin{itemize}
    \item In low-density networks, where $H$ is small, $\varrho_d$ is larger, enabling clusters to form over greater distances and connect sparse nodes.
    \item In high-density networks, $\varrho_d$ is smaller, preventing clusters from becoming too large and maintaining communication efficiency.
\end{itemize}
Through this function, the clustering will be dynamic with changes in the density of nodes. This necessitates determining the best value for the parameter $\Delta$ in the Eq. \ref{Eq_15:ADT}, which depends on various factors. These factors include the Average Cluster Size (ACS), which helps maintain clusters of an optimal size by balancing between clusters that are too large or too small. It is obtained by dividing the total number of participating nodes by the total number of clusters. This metric provides insights into the effectiveness of the clustering mechanism and its ability to maintain well-structured group formations. Moreover, we take into account the Intra-Cluster Communication Cost (ICCC), which measures the communication expenses within a cluster. For any cluster $C_i$ the ICCC is determined by summing the weights of the edges that link nodes within $C_i$. A lower ICCC signifies more efficient communication among nodes within the same cluster, resulting in reduced costs. This is advantageous because it ensures that related or frequently interacting nodes are grouped together, thereby minimizing communication overhead within clusters. The Inter-Cluster Communication Cost (ECCC), on the other hand, quantifies the communication expenses between different clusters. For clusters $C_i$ and $C_j$ (where $i\neq j$), the ECCC is calculated as the sum of the weights of the edges connecting nodes between $C_i$ and $C_j$. A lower ECCC indicates that minimal communication is needed between nodes in separate clusters, which is generally preferable since cross-cluster communication often leads to higher latency, greater resource consumption, and potential bottlenecks. Another important factor is the Message Delivery Success Rate (MDSR), which measures the percentage of messages that are successfully delivered to their intended destination out of the total number of messages generated. This metric is a key performance indicator for evaluating the reliability of a routing protocol, where higher values indicate improved message delivery efficiency and lower packet loss. In our method, we achieved the best value for $\Delta$, after different simulation studies. For example, after running different simulations with 500 nodes, we obtained the results shown in Table \ref{tab_3}.
\begin{table}[!ht]
\centering
\caption{The best value of $\Delta$}
\label{tab_3}
\begin{tabular}{| p{0.5cm} | p{1.5cm} | p{1.5cm} |p{1.5cm}|p{1.5cm}|}
\hline
\textbf{$\Delta$} &\textbf{ACS} & \textbf{ICCC}&\textbf{ECCC}&\textbf{MDSR} \\
\hline
0.5& Small & Low&High&Moderate\\
\hline
1&Moderate&Moderate&Moderate&High\\ 
\hline
1.5& Moderate&Low&Low&High\\
\hline 
2&Very Large&Very High&Very Low&Low \\
\hline
\end{tabular}
\end{table}
Our simulations indicate that a scaling factor of $\Delta=1.5$ achieves the best balance between cluster size, intra-cluster communication cost (ICCC), inter-cluster communication cost (ECCC), and message delivery success rate (MDSR). A lower $\Delta$ results in small clusters with high inter-cluster communication costs, whereas a higher $\Delta$ leads to oversized clusters with inefficient intra-cluster communication. The chosen value minimizes both communication overhead and cluster instability.\\
Assuming a network with 500 nodes ($N$) distributed over an area ($U$) of 1000 square units, therefore, $H$ can be calculated as: 
\begin{equation} \label{Eq_25:HExam}
    \centering
    H=\frac{500}{1000}= 0.5 \ nodes \ per \ {{unit}^2}
\end{equation}
And an average Manhattan distance:
\begin{equation} \label{Eq_26:DExam}
    \centering
    d \approx \frac{1}{\sqrt{0.5}} \approx 1.414 \ units
\end{equation}
Finally, we can obtain the ADT ($\varrho_d$):
\begin{equation}\label{Eq_27:ADTExam`}
    \centering
    \varrho_d= 1.5*1.414 \approx 2.121 \ units
\end{equation}
After clustering, routing begins, a novel routing approach is proposed to enhance communication efficiency between the source (${S}$) and destination (${D}$). This routing technique emphasizes inter-cluster communication, facilitating interaction among different clusters (${C}_{i}\cap {C}_{j}\neq0$). This enables the optimal selection of subsequent relay and forwarding nodes towards the destination. To enable seamless integration with TCP/IP, CR-DRL utilizes a middleware layer that represents the dynamic clusters within the OppNets as virtual network segments. A virtual network segment is a logically defined portion of the network where nodes dynamically organize themselves based on connectivity and communication potential, rather than being constrained by fixed physical infrastructure. These segments are continuously updated as nodes move, allowing for efficient data transmission even in highly dynamic environments. Within this structure, cluster heads act as gateway nodes, managing logical paths that serve as temporary yet stable communication links. These links are referred to as virtual links because they are not permanently established physical connections but rather dynamically formed logical routes that adapt to changing network conditions, node availability, and routing requirements. Unlike traditional wired or wireless connections, these virtual links exist only while communication opportunities are available, making them well-suited to the opportunistic nature of the network. By leveraging these virtual links, standard TCP/IP protocols can be effectively implemented, ensuring reliable routing and data delivery even in highly dynamic conditions. Additionally, CR-DRL employs shared nodes as communication relays to enhance the consistency of message forwarding, thereby improving compatibility with TCP/IP transmission requirements. This approach helps mitigate the effects of intermittent connectivity by preserving logical communication paths, ensuring a more stable and continuous data flow across multiple hops. Figure \ref{fig_2:Routing} depicts the system model, where vehicles are organized into clusters based on their proximity and communication range, with cluster heads overseeing data transmission. Shared nodes play a key role in enabling inter-cluster communication, ensuring seamless network connectivity. 
In highly dynamic OppNets, packet loops may arise when multiple common members forward the same message to different clusters simultaneously, creating redundant paths that could route the packet back to its original cluster. To mitigate this, CR-DRL selects the most suitable common member based on the shortest Manhattan distance, minimizing transmission distance and avoiding unnecessary forwarding. Additionally, each message retains a record of previously visited clusters to ensure it is not retransmitted to the same location. A Time-To-Live (TTL) constraint of 300 seconds discards undelivered packets after a set duration, while a hop count limit of 10 restricts excessive forwarding between clusters. These measures collectively enhance routing efficiency and effectively prevent packet looping, even in highly dynamic network environments.
\begin{comment}
   For instance, let's consider the scenario where the source node  ${v}_{5}$ within cluster ${C}_{1}$ intends to transmit a packet destined for node ${v}_{11}$ in cluster ${C}_{3}$. Initially, ${v}_{5}$ dispatches a message to its cluster head ${\chi}_{1}$. Based on the destination address (${v}_{11}$), ${\chi}_{1}$ checks its list of shared members. If the destination (${v}_{11}$) is found in its list, the packet is directed straight to ${v}_{11}$. However, if ${v}_{11}$ is not its list, ${\chi}_{1}$ forwards the packet to the common node ${\aleph}_{1}$, which serves as the intermediary between clusters ${C}_{1}$ and ${C}_{2}$. Subsequently, relays the packet to its cluster head node ${\chi}_{2}$. Should the destination node be included in its table of ${\chi}_{2}$, forwards the packet directly to the destination ${v}_{11}$. If not, the packet is routed to the common node ${\aleph}_{2}$ in other clusters until it reaches its intended destination.\\  Using this method, the packet is routed reliably to its destination.
\end{comment}
As a final step, the clustering process is repeated after a time interval. The high mobility of nodes in these networks makes it possible for any node to leave a cluster and for new nodes to join it. 
The proposed CR-DRL algorithm (Algorithm \ref{Algo_1:CR-DRL}) facilitates efficient message routing in OppNets by leveraging DRL. When a vehicle node $v_i$ sends a message to a distant node $v_j$, GPS technology provides location data, and static parameters such as vehicle count ($ N $), area ($ U $), adaptive threshold factor ($ \Delta $), learning rates ($ \alpha_a $, $ \alpha_C $), and discount factor ($ \gamma $) are set. The algorithm calculates Manhattan distances between vehicles (Line 1-4), determines maximum encounter rate, remaining energy, and buffer size using specified equations (\ref{Eq_17:maxER}, \ref{Eq_18:maxRE}, and \ref{Eq_19:BS}), and defines the state ($ s $) with factors ${\phi_{{v}_{i}}}_{\max}$, $ \Psi_{v_i} $, and $ \eta_{v_i} $ (Line 6-8). A reward function based on delivery ratio, throughput, and delay guides the AC updates to select optimal cluster heads, which are stored in a ClusterHead array (Line 9-17). Using the values of $N$ and $U$, the node density is first determined. The ADT, ($ \varrho_d $) determine cluster size ($ \beta_{C_i} $) (Line 18-19). In the routing phase, the source node sends the message to its cluster head ($ \chi_i $), which delivers it directly if the destination is a member or forwards it to another cluster head ($ \aleph_q $) until reaching $ v_j $ (Line 20-27). Key aspects include dynamic cluster head selection, adaptive clustering via ADT, and efficient routing to optimize network performance.
\begin{algorithm}[!t] \caption{CR-DRL}\label{Algo_1:CR-DRL}  
 \hspace*
\algorithmicindent\textbf{Input:~}GPS data, ClusterHead Array, $N$, $U$, $\Delta$, $\gamma$, $\alpha_a$, $\alpha_C$
\begin{algorithmic}[1]
    \For {${v}_{i}\in {v}$}\label{loop1}
    \For {${v}_{j}\in {v}$}\label{loop2}
\State ${d}={|{x}_{i}-{x}_{j}|}+{|{y}_{i}-{y}_{j}|}$
\EndFor\label{loop2}
\EndFor\label{loop1}
\State Calculate $\phi_{{v_i}_{max}},\Psi_{v_i},\eta_{v_i}$
\For {each episode}
\State Initial state $s=\{\phi_{{v_i}_{max}},\Psi_{v_i},\eta_{v_i}\}$
\For {each time step $t$ in  the episode}
\State Select action $a$ (select cluster head) using policy
\State  Calculate reward $R$ using Equation \ref{Eq3_Reward}
\State Compute TD error using Equation \ref{Eq8_TD} 
\State Critic update using Equation \ref{Eq9_CriticU}
\State Actor update using Equation \ref{Eq7_ActorU}
\EndFor
\EndFor
\State Sorted from maximum to minimum cluster heads (${\chi_1},{\chi_2},{\chi_3},...,{\chi_n}$) , ClusterHead Array  is updated
\State Calculate $H$, $\varrho_d$
\State $\beta_{C_i}$=  $\varrho_d$
\State ${v}_{i}\rightarrow$ Send packet to ${\chi}_{i}$
\State ${\chi}_{i}\leftarrow$ Receive the Packet
\State ${\chi}_{i}$ check its list
\If{the destination address is at ${\chi}_{i}$ list} \label{Checklist}
\State ${\chi}_{i}\rightarrow$ send the packet to the 
destination
\State ${v}_{i}\leftarrow$ Receive the packet through it ${\chi}_{i}$
\Else
\State
 ${\chi}_{i}\rightarrow$ Send packet to Common Member $\aleph_{q}$ in ${C}_{i}$  
\State Repeat lines 20 to 27   \EndIf\label{Checklist}
\State interval time=0
\State Repeat line 1
\end{algorithmic}
       \end{algorithm}
       \vspace{-6pt}
\subsection{Time complexity}
In terms of computational complexity, the AC algorithm (${T}_{AC}$) updates its policy parameters iteratively over $m$ training steps, adjusting $E$ neural network weights at each step. Thus, its complexity is given by $O(mE)$, capturing the dependency on training iterations and model size. Clustering involves assigning each of the $X$  nodes to one of $C$ clusters based on encounter rate, energy, and buffer size. Each node's cluster membership requires a comparison with other nodes, leading to a time complexity of $O(XC)$. Routing involves forwarding messages between clusters using shared members ($M$), which act as relay nodes. Since each cluster ($C$) must communicate through its shared members, the routing complexity is given by $O(CM)$, capturing the interactions between clusters and common nodes. 
Consequently, the total computational time for CR-DRL can be summarized as follows:
\begin{equation}\label{Eq_28:ComputinalTime}
    \centering
    {T}_{Total}=O(mE)+O(XC)+O(CM)
\end{equation}
The overall computational complexity of CR-DRL grows with the number of training iterations ($m$),  the neural network size ($E$), the number of nodes ($X$), and the number of clusters ($C$). In practical deployments, $X$ and $C$ are expected to be significantly smaller than $mE$, meaning the dominant factor in computational cost is reinforcement learning training. However, optimizing the number of clusters and minimizing unnecessary shared members ($M$) can significantly improve routing efficiency and reduce overhead.
\section{Simulation setting and evaluation metrics}\label{simulation study}
\begin{figure}[!t]
    \centering
    \subfloat[Policy and Value loss convergence.\label{PVL}]{
        \includegraphics[width=2.5 in]{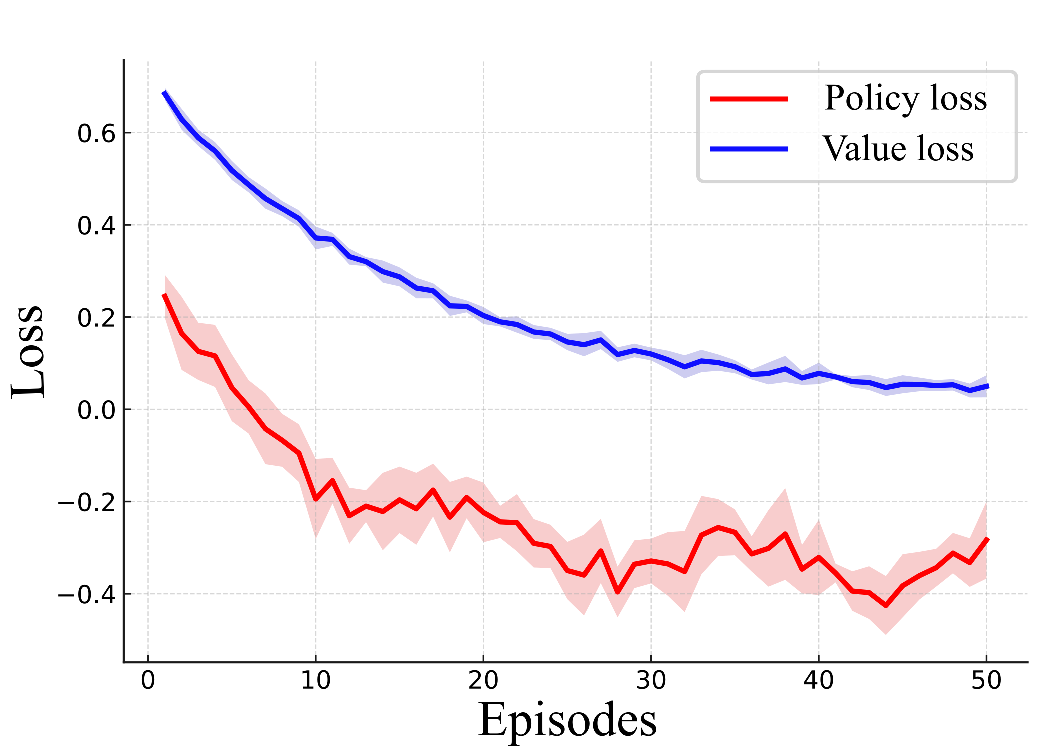}
    }\\
    \subfloat[Cumulative rewards.\label{CUReward}]{
        \includegraphics[width=2.5 in]{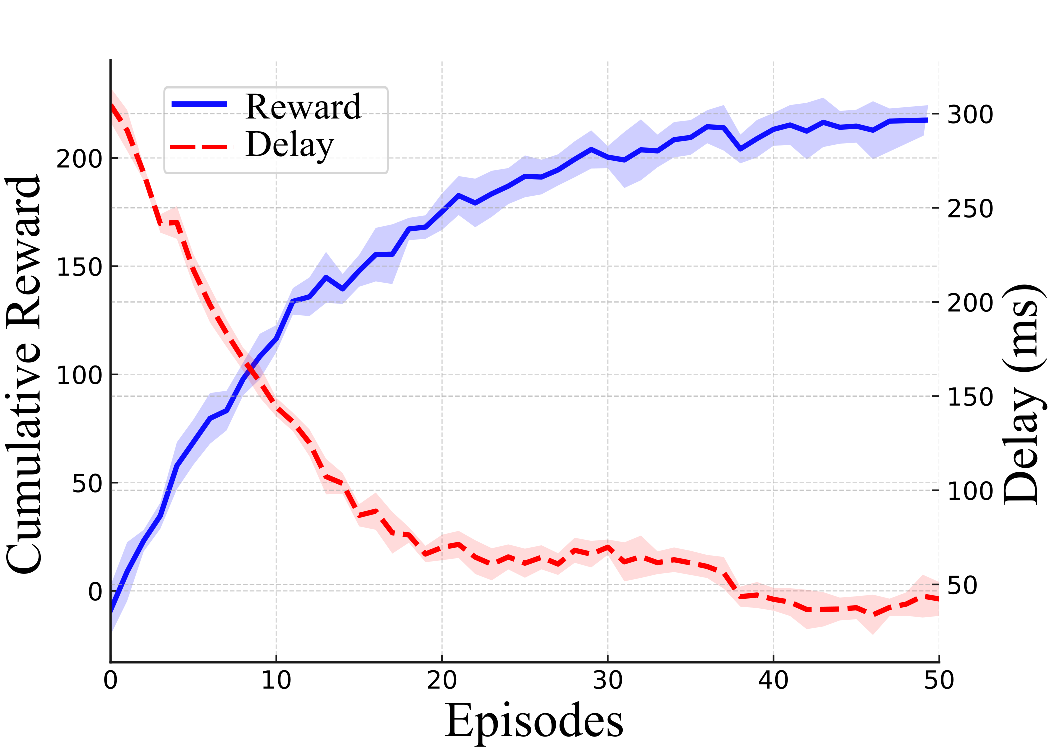}
    }
    \caption{Convergence of policy and value loss (a), and cumulative reward (b) in the AC learning framework.}
    \label{AClearning2}
    \vspace{-0.1in}
\end{figure}

This simulation study evaluates the performance of the proposed CR-DRL protocol in dynamic OppNets. Specifically, we analyze its ability to form stable clusters and maintain efficient data routing under varying network conditions. We evaluate CR-DRL's performance in relation to mobility patterns and vehicle movement using the Opportunistic Network Environment (ONE) simulator. ONE is an open-source Java-based simulation tool for developing and testing routing protocols \cite{A34}. 
With ONE, it is easy to integrate contact records, route modules, applications, and reports. For the CR-DRL, we have used the default ONE simulation map from Helsinki, Finland. The geographical space for this experiment is ${4000*3500m}^{2}$ in size, with 50, 100, 150, 200, and 250 nodes distributed at random. The GPS coordinates, defined by longitude and latitude, are based on geographical space. To reflect real-world heterogeneous mobility, we use a mixed mobility model where 50\% of nodes follow a Random Waypoint Model (RWP), 30\% follow a Gauss-Markov pattern with dynamic speed and direction changes, and 20\% exhibit clustered movements based on the Working Day Movement Model (WDM) model. This setup ensures diverse and realistic node mobility, accounting for varying speeds, path choices, and encounter frequencies \cite{A35}. Each message generated within the network has a TTL of 300 seconds and varies in size from 250 to 750 KB. The nodes have buffer capacities between 50 and 100 MB. To simulate both low and high traffic loads, the message generation interval per node was randomly adjusted between 5 and 40 seconds. To emulate realistic heterogeneous mobility, vehicles were categorized into two groups: 80\% cars with speeds between 8–15 m/s and 20\% service vehicles (e.g., trams) with speeds of 4–8 m/s. A failure model was incorporated where 5\% of nodes were randomly deactivated between times 1800–2400 seconds to simulate vehicle dropouts or hardware failures. The simulation parameters are set for a duration of 1 hour, with an initial energy of 4800 Joules, scanning energy of 0.92 mW/s, transmission energy of 0.08 mW/s, receive energy of 0.08 mW/s, an energy recharging interval of 2800 seconds, common members transmit range 100M and cluster head transmit range 300M. As mentioned in the section \ref{System model}, if a vehicle is unable to communicate with its cluster head due to being out of range, it momentarily increases its transmission range to restore connectivity. Initial experiments indicated that a 20\% expansion in transmission range is typically sufficient to maintain communication in most scenarios. However, since this adjustment occurs only when required and has a minimal impact on overall network performance, it was excluded from the primary simulations to maintain focus on key metrics such as clustering efficiency, energy consumption, and routing performance. The simulation employs the ONE simulator’s standardized mobility model derived from the Helsinki map. Figure \ref{AClearning2} presents the convergence characteristics of the CR-DRL framework during the learning process.
Figure \ref{PVL} presents the evolution of the policy loss and value loss across 50 training episodes. An episode denotes a complete routing cycle from the initial state to the terminal condition, either successful delivery or timeout. The policy loss, derived from the entropy-regularized objective in Eq. \ref{ANFunction}, shows a decreasing trend with moderate fluctuations due to the stochastic nature of policy gradient updates and the exploration term $\varepsilon O(\pi_{\theta})$. Despite these oscillations, the loss converges within 50 episodes, confirming that the actor progressively improves action selection. In parallel, the value loss measures the critic's error in estimating state values $V(s;\omega)$, showing the critic's improving ability to predict expected returns as it converges toward lower values. Together, these results validate that the AC algorithm achieves stable and efficient convergence under the dynamic vehicular environment. 

Figure \ref{CUReward} illustrates the relationship between cumulative reward and average delay over 50 episodes. The cumulative reward, computed from the structure in Eq. \ref{Eq3_Reward}, increases steadily from low initial values and converges near 200, while the average delay decreases from approximately 280 ms to 50 ms as training progresses. This inverse correlation confirms that the reward function effectively promotes higher delivery ratio and throughput while reducing delay, thereby demonstrating the stability and robustness of the proposed framework and the efficacy of its learning dynamics. 

The definition of key metrics such as delivery ratio, throughput, delay is as follows: The delivery ratio is to measure the proportion of data packets successfully delivered to their destination out of the total sent \cite{A36}. E2E delay measures the time a packet takes to travel from source to destination across a network \cite{A37}. A hop count indicates the total number of intermediate devices, such as nodes, that a data packet passes through from its source to its destination. Each node along the data path counts as a hop that facilitates the transfer of data from one point to another \cite{A38}. Throughput indicates the volume of data in bits transferred from a source to a destination per unit of time (seconds) over a communication link \cite{A39}. Residual energy represents the remaining battery capacity of each node, calculated as the initial energy minus the cumulative consumption from operations such as scanning, transmission, and reception. Normalized performance scales key metrics including delivery ratio, delay, and energy consumption between 0 and 1 to enable fair comparison \cite{A36}. Convergence rate quantifies the speed and stability of policy improvement in the RL algorithm, assessed through reward stabilization and loss reduction over successive training epochs.
\vspace{-6pt}
\subsection{Parameter analysis}
\begin{figure*}[!t]
    \centering
    % Top row
    \subfloat[Sensitivity of delivery ratio to AC learning rates.\label{ACLearning}]{
        \includegraphics[width=2.5 in]{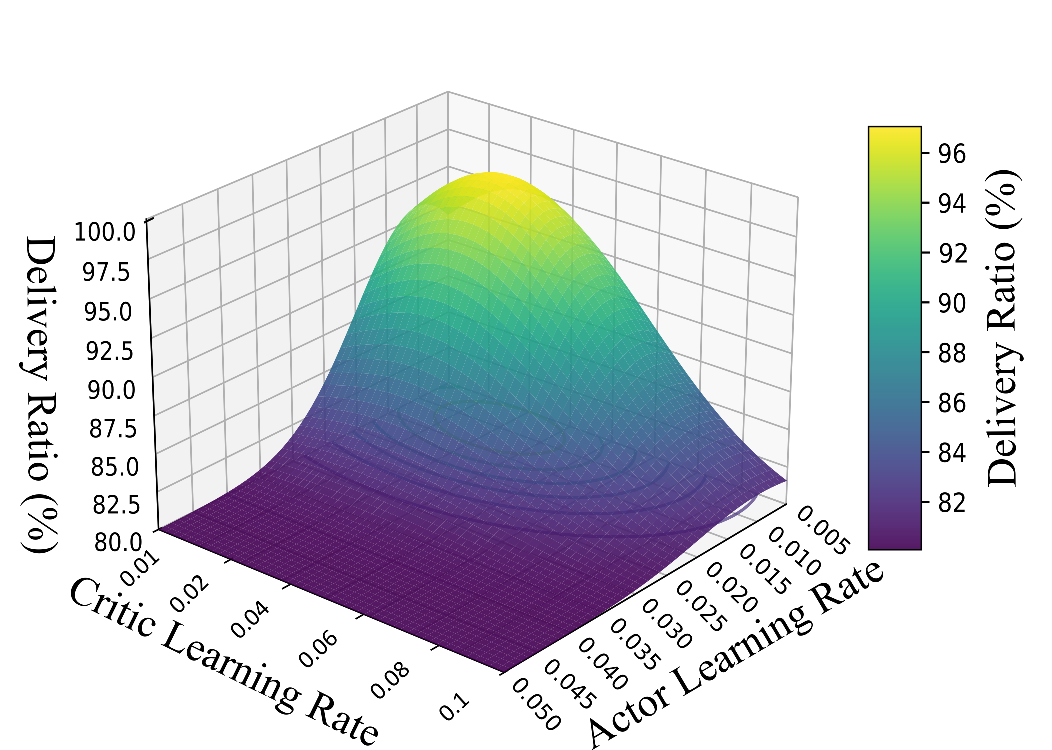}
    }
    \hspace{0.04\linewidth}
    \subfloat[Effect of discount factor $\gamma$.\label{DFactor}]{
        \includegraphics[width=2.5 in]{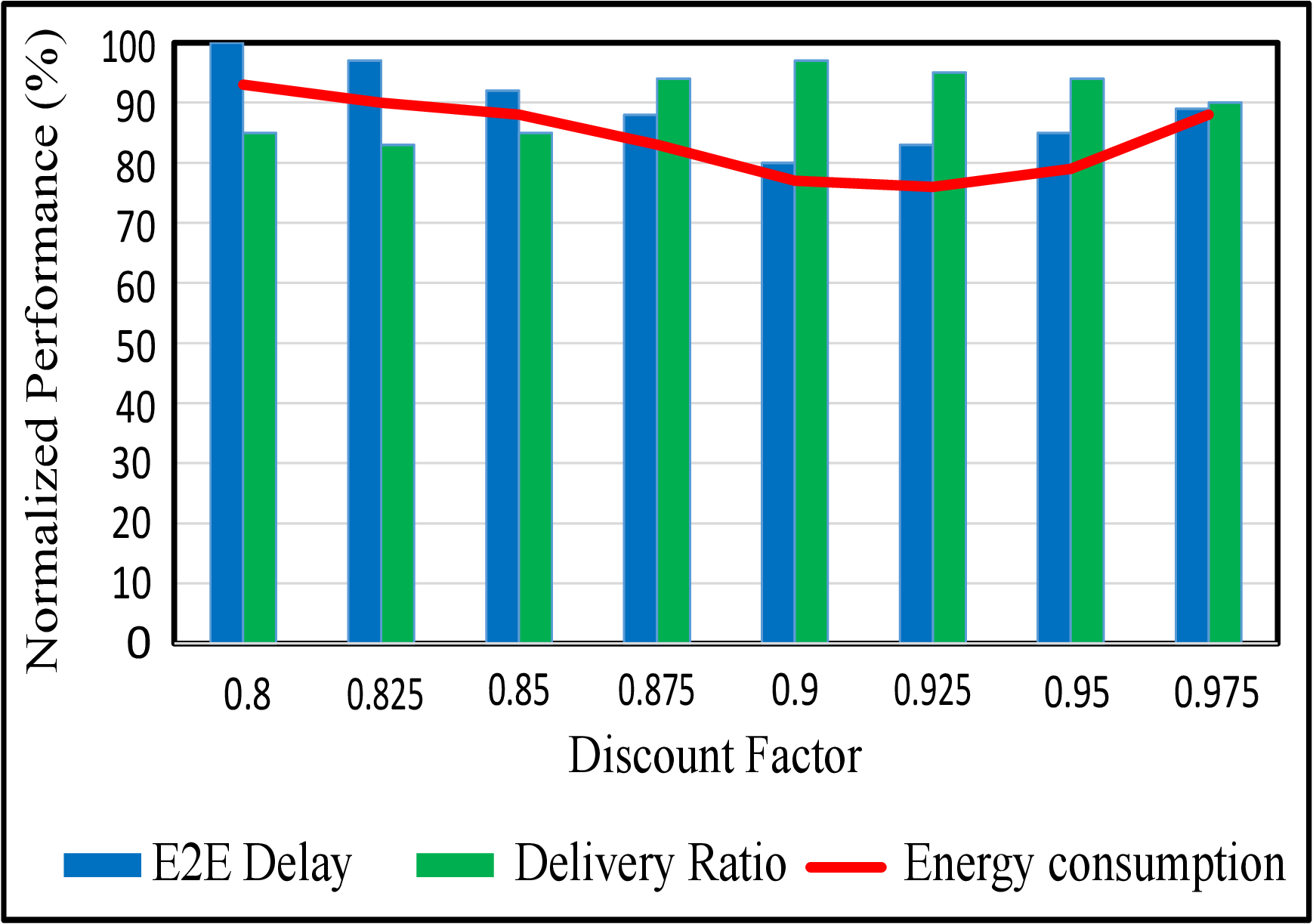}
    }
    \\[1ex]
    % Bottom row
    \subfloat[Impact of $\Delta$ on clustering efficiency.\label{Delta}]{
        \includegraphics[width=2.5 in]{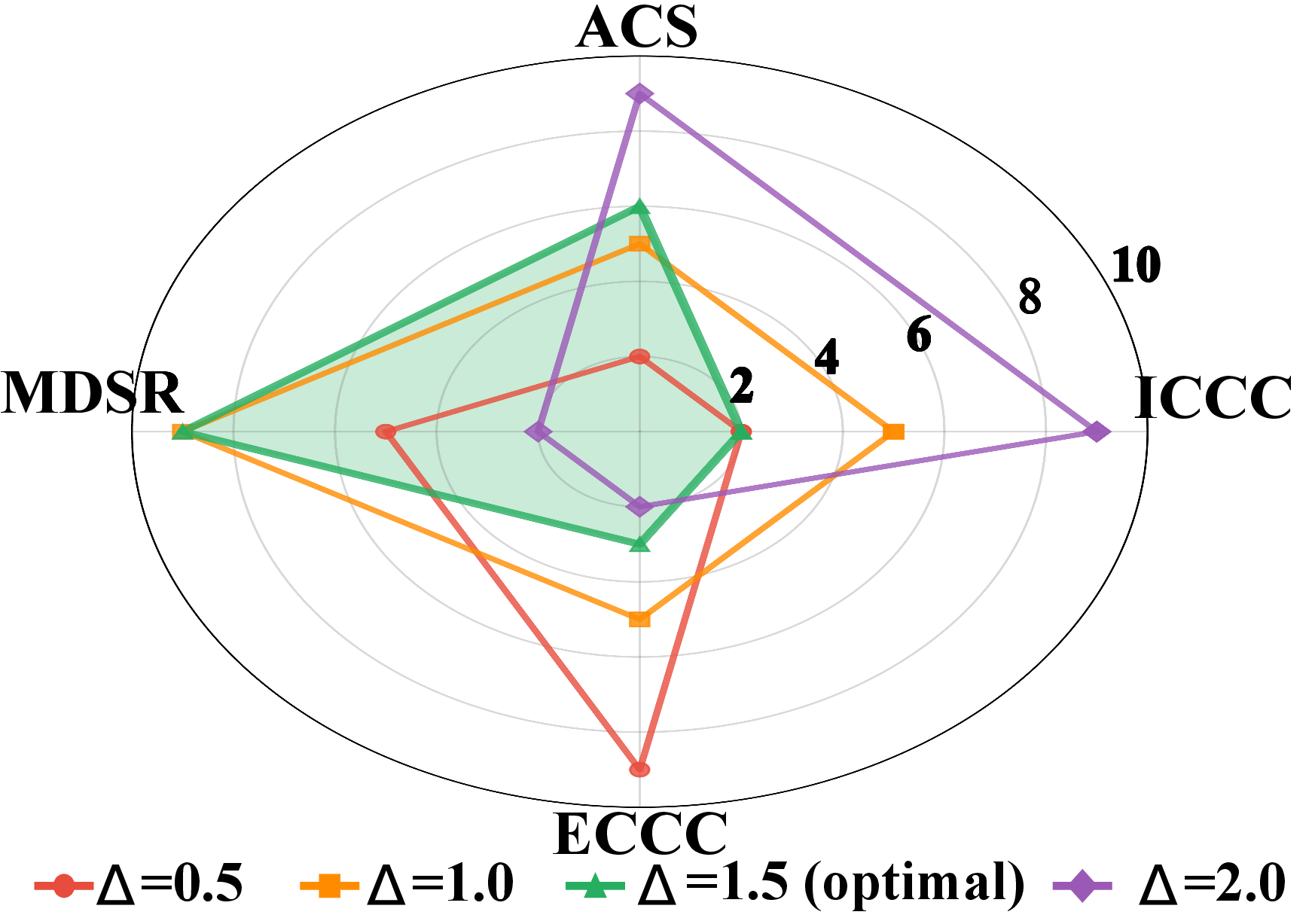}
    }
    \hspace{0.04\linewidth}
    \subfloat[Influence of entropy coefficient on exploration–exploitation balance.\label{entropy}]{
        \includegraphics[width=2.5 in]{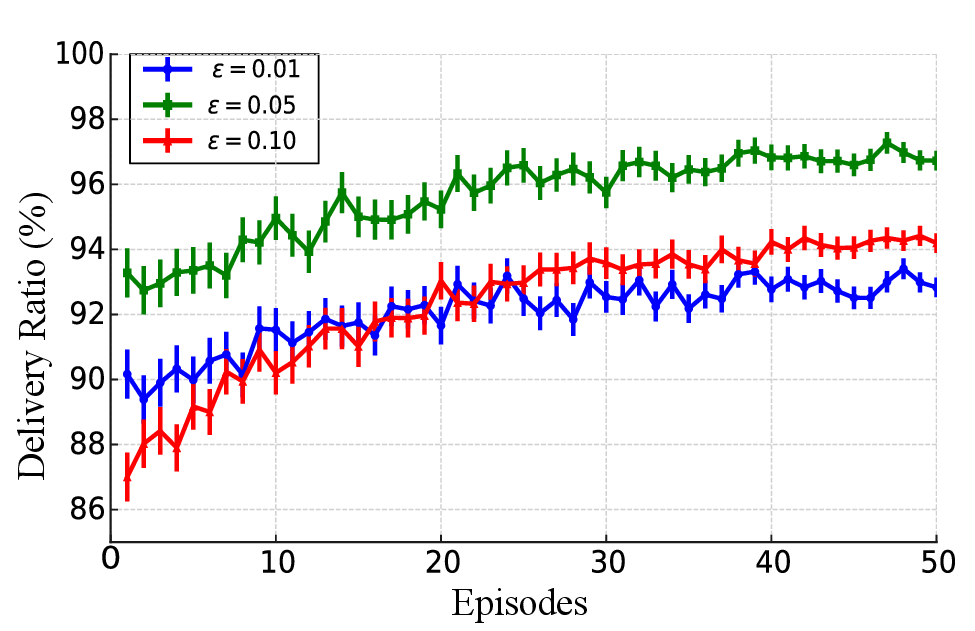}
    }
    \caption{Sensitivity analysis of CR-DRL hyperparameters: (a) impact of AC learning rates on delivery ratio; (b) effect of discount factor on normalized performance; (c) influence of distance scaling factor on clustering efficiency metrics; (d) role of entropy coefficient in balancing exploration and convergence.}
    \label{AllMetricsC}
    \vspace{-0.1in}
\end{figure*}

We investigate the influence of key AC hyperparameters including $\gamma$, $\alpha_a$, $\alpha_C$, $\varepsilon$, and $\Delta$ on the performance of CR-DRL. To ensure statistical reliability and mitigate the effects of random variation, each experimental setup was independently executed across 50 episodes. The parameter $\alpha_a$ was systematically varied over the interval $[0.005, 0.5]$, whereas $\alpha_C$ was examined within the range $[0.01, 0.1]$. The combined impact of these hyperparameters on delivery ratio is depicted in Fig. \ref{ACLearning}. Results indicate that the setting $\alpha_a = 0.01$ and $\alpha_C = 0.02$ provides the most consistent convergence behavior, attaining delivery ratios as high as 97\%. Additionally, Fig. \ref{DFactor} demonstrates the influence of $\gamma$, evaluated across the interval $[0.8, 0.99]$, on normalized performance. A value of $\gamma = 0.9$ provides the best trade-off, maintaining a high delivery ratio while controlling delay and energy consumption. Values of $\gamma$ exceeding 0.95, as well as those below 0.85, resulted in diminished delivery performance, increased latency, and elevated energy consumption. $\varepsilon$ was varied within $[0.01, 0.1]$, and $\Delta$ within $[0.5, 2.0]$ (Figures \ref{Delta} and \ref{entropy}).
The selected values, $\varepsilon = 0.05$ and $\Delta = 1.5$, strike an effective trade-off between exploration and convergence. In particular, $\varepsilon = 0.05$ achieved up to 97\% of the optimal delivery ratio within 50 episodes while ensuring stable convergence. At extreme values of $\Delta$, CR-DRL performance degrades significantly. When $\Delta = 2.0$, the clustering process generates excessively large clusters, resulting in elevated ICCC and a marked reduction in MDSR. In contrast, setting $\Delta = 0.5$ excessively fragments the network into small clusters, which increases the ECCC and yields only moderate performance improvements. As summarized in Table \ref{tab_3}, an intermediate configuration of $\Delta = 1.5$ achieves the most effective trade-off, delivering a balanced ACS, minimized ICCC and ECCC, and consistently high MDSR.
\subsection{First scenario}
\begin{figure}[!t]
    \centering
    \subfloat[Comparison of vehicles and clusters using the ADT function.\label{ClusterandVehicle}]{
        \includegraphics[width=2.5 in]{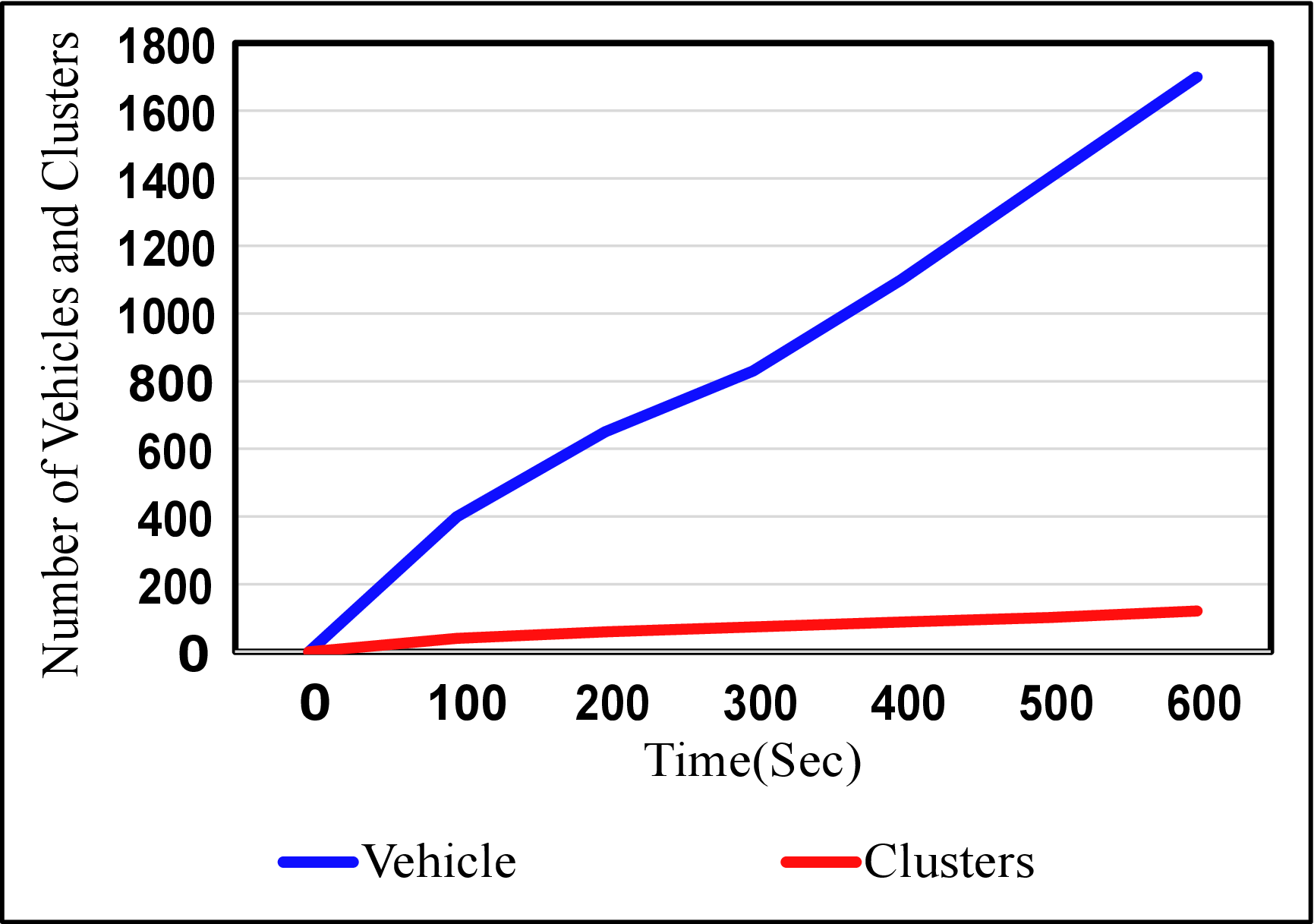}
    }\\
    \subfloat[Comparison of vehicles and clusters without application of the ADT function.\label{ClusterandVehicle1}]{
        \includegraphics[width=2.5 in]{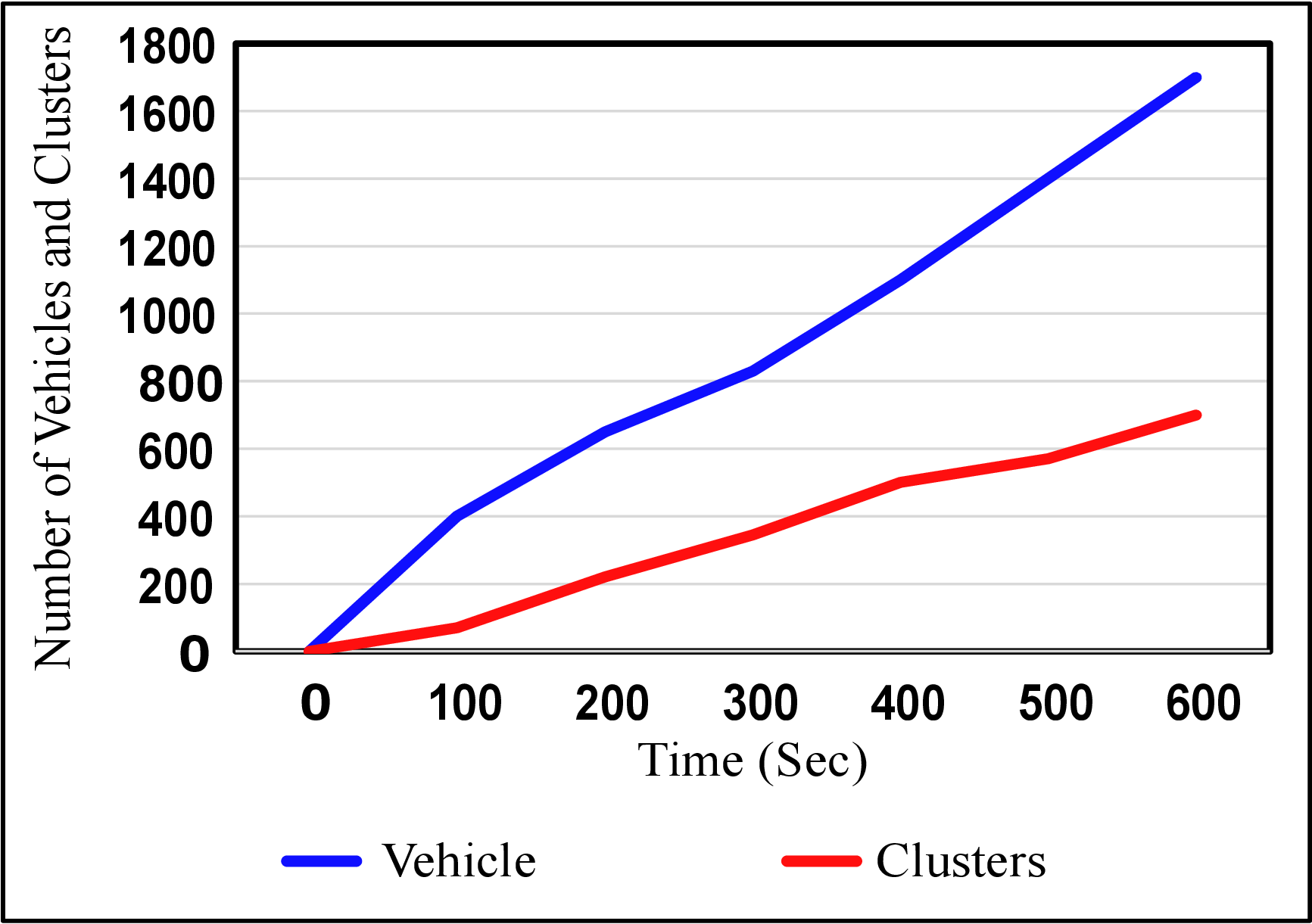}
    }
    \caption{Vehicles vs. clusters with ADT (a), and (b) without ADT.}
    \label{VehicleandClusterGeneral}
\end{figure}

\begin{figure}[!t]
    \centering\includegraphics[width=2.5in]{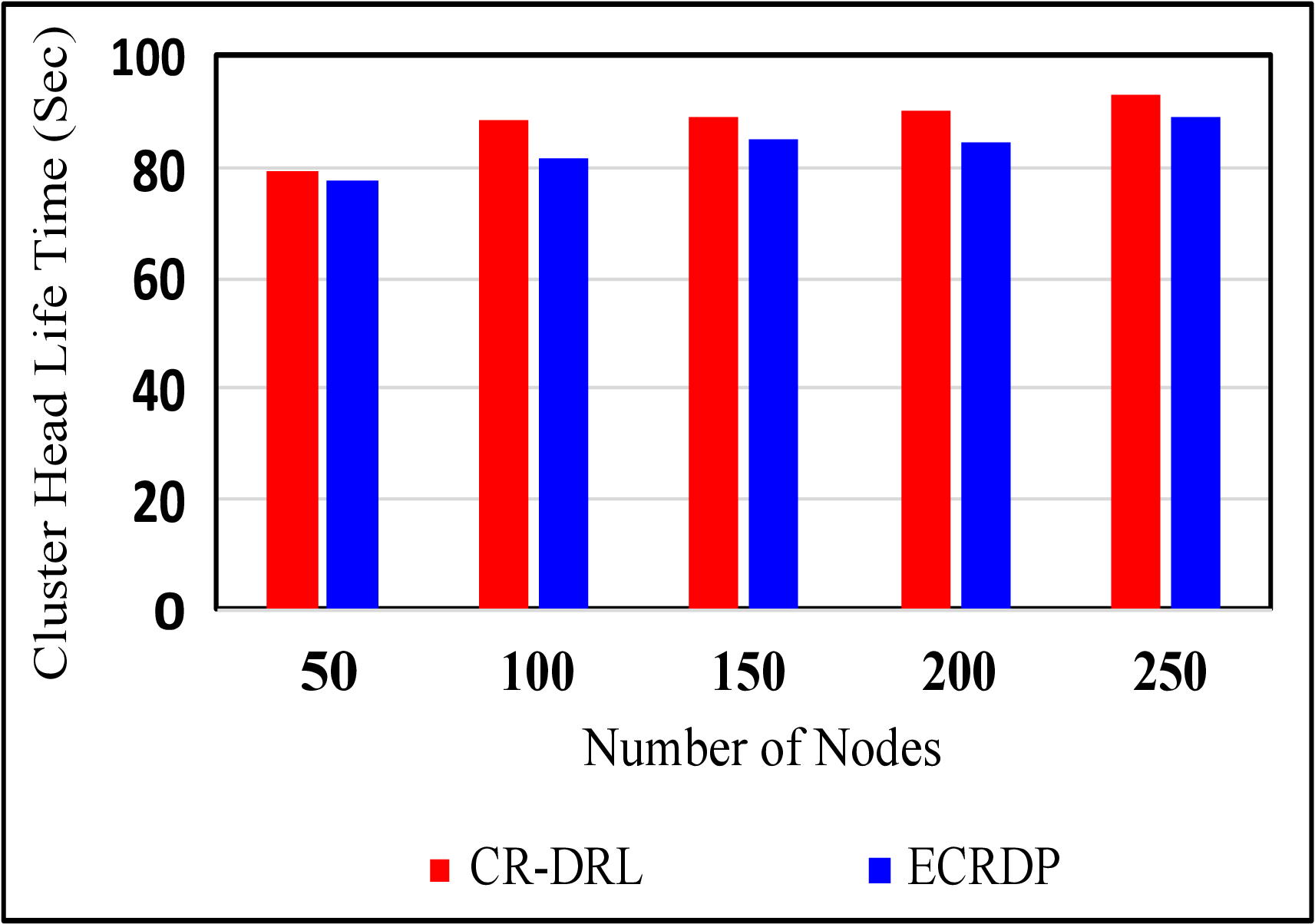}
    \caption{Cluster head lifetime in CR-DRL method versus ECRDP method.}
    \label{Fig_6:CHL}    
\end{figure}

\begin{figure}[!t]
    \centering\includegraphics[width=2.5in]{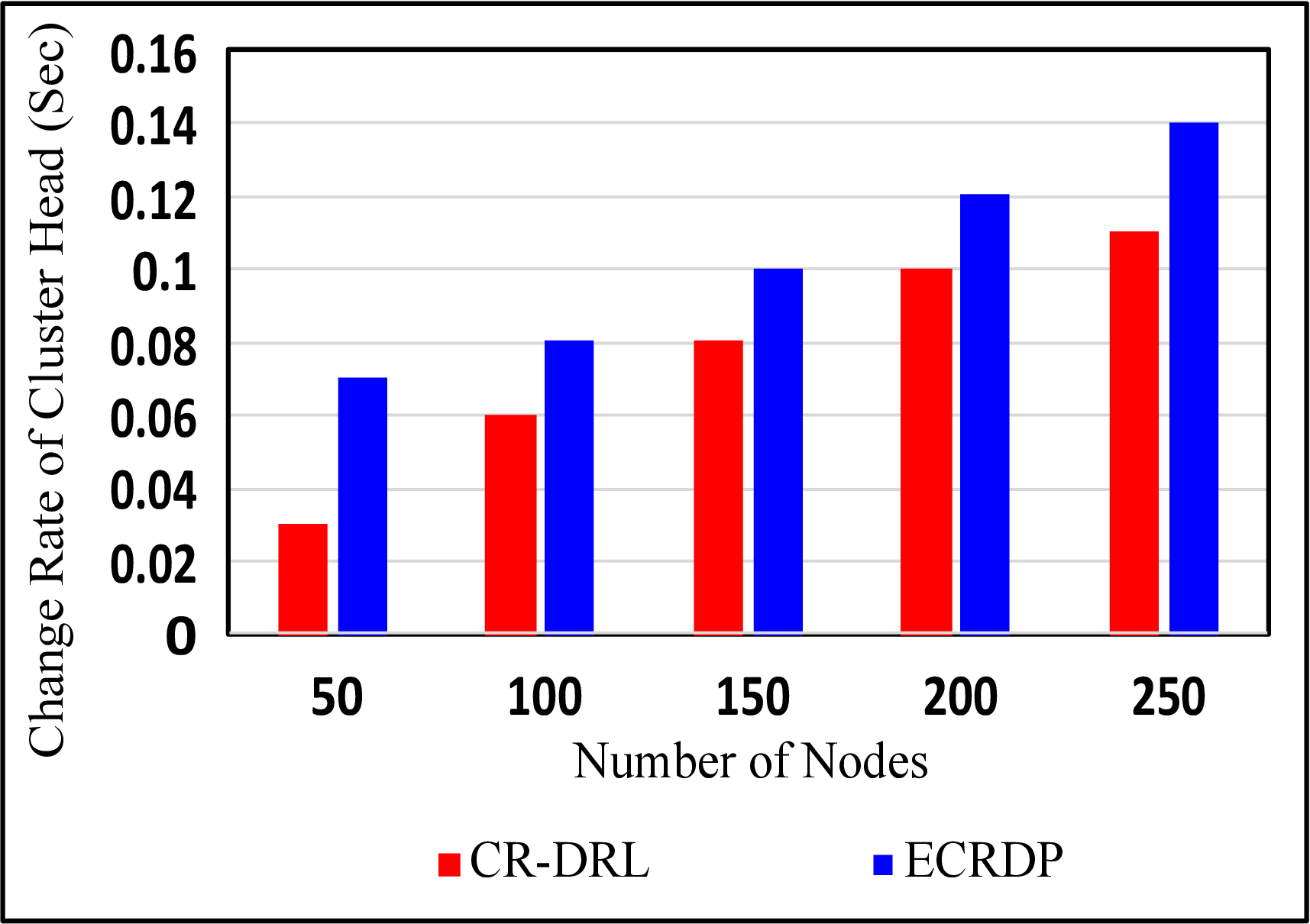}
    \caption{Cluster head change rate in CR-DRL method versus ECRDP method.}
    \label{Fig_7:CHCR}    
\end{figure}

\begin{figure}[!t]
    \centering
    \includegraphics[width=2.5 in]{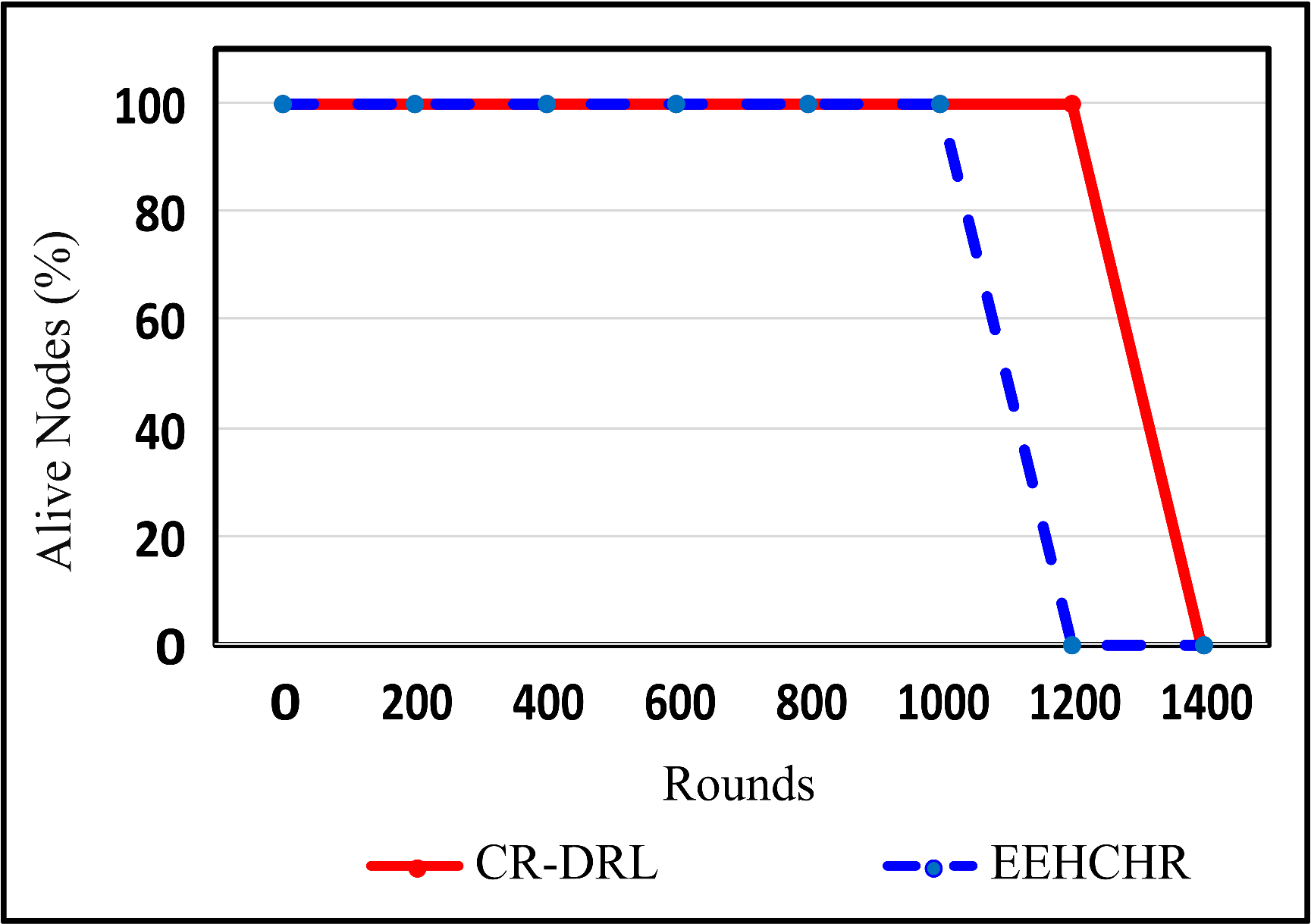}
    \caption{CR-DRL and EEHCHR network lifetime analysis.}
    \label{FigNew1}
\end{figure}

In the first scenario, a variety of performance criteria are taken into account to ensure a more comprehensive and objective analysis. These criteria include the extent of overlap between clusters, an evaluation of the proposed dynamic clustering defined by Eq. \ref{Eq_15:ADT}, and the stability of both the cluster heads and the cluster members and the overall network lifetime.
In our simulation study, we evaluated the average overlapping degree both with and without applying this adjustment function. It is important to note that our method does not always guarantee a fully connected communication path meaning the total connectivity ratio $\tau_{t}(G)$ may not reach 100\%, and there may be instances where no communication path exists between the source and the destination. To accurately evaluate the effectiveness of Eq. \ref{Eq_15:ADT}, we started with a small number of vehicles and gradually increased the number to a maximum of 1800. Additionally, during a maximum observation period of 600 seconds, we analyzed Eq. \ref{Eq_15:ADT} to determine whether this relationship has a significant impact on the clustering process and the overall network performance. Figures \ref{VehicleandClusterGeneral} shows the effects of using Eq. \ref{Eq_15:ADT}. In Figure \ref{ClusterandVehicle1}, an increase in the number of vehicles corresponds with a rise in the number of clusters. This could be due to the formation of many clusters that do not use the ADT function and clustering is static.
Figure \ref{ClusterandVehicle} illustrates the correlation between the quantity of vehicles and the count of clusters through Eq. \ref{Eq_15:ADT}. Over time, the red line representing the number of clusters appears consistent, indicating a stabilization in the CR-DRL topology. This stability is crucial in real-world OppNets, where frequent re-clustering can result additional overhead and delay packet forwarding. CR-DRL’s ability to maintain balanced clusters even as node density fluctuates ensures efficient and reliable communication across dynamic urban traffic environments. In dense environments, ADT restricts cluster expansion to avoid excessive overlap, while in sparse networks, it allows clusters to span wider areas, maintaining connectivity. This mechanism ensures an optimal balance between stability and routing efficiency, preventing excessive re-clustering while maintaining strong inter-cluster links.
\begin{figure}[!t]
    \centering
\includegraphics[width=2.5in]{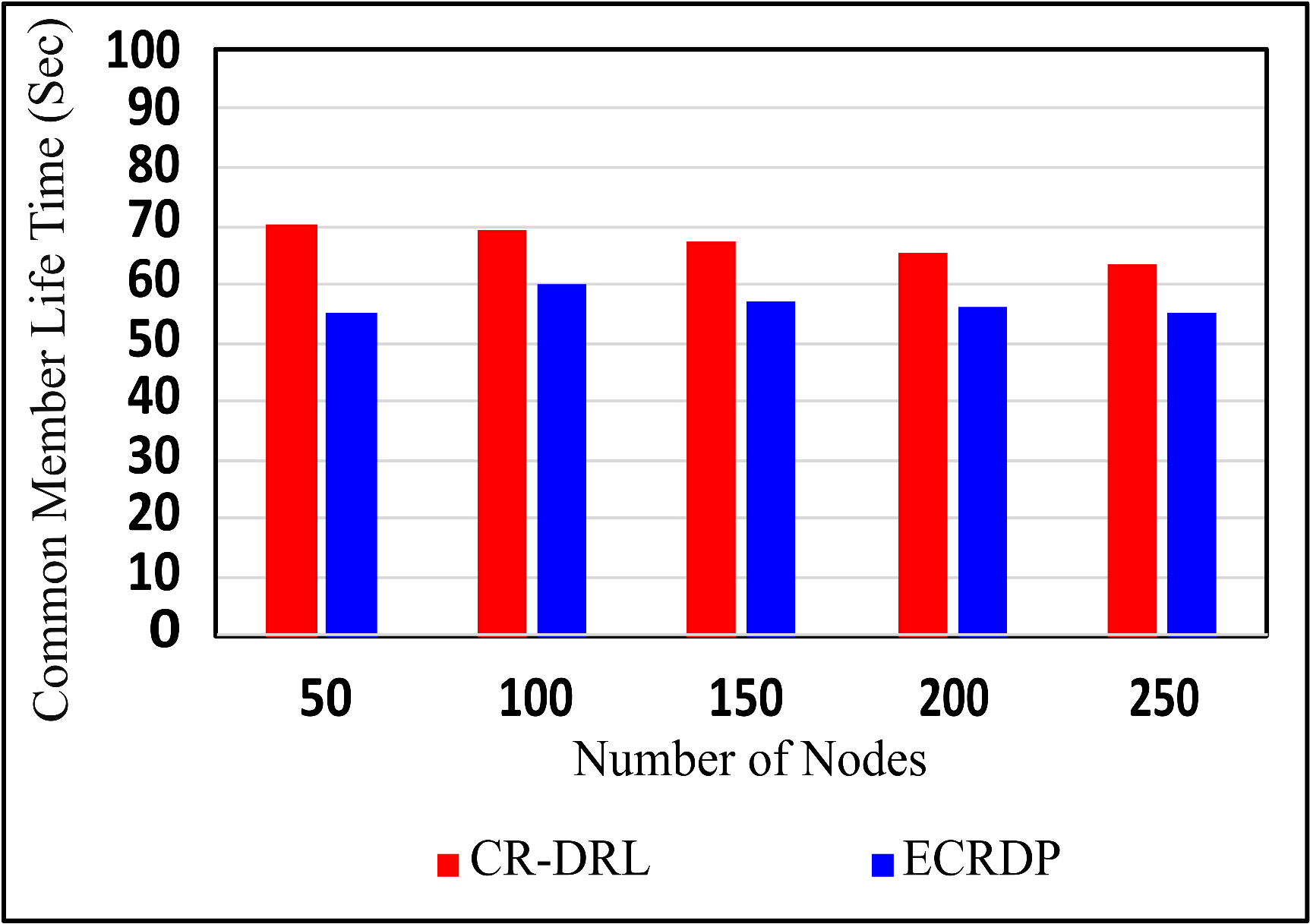}
    \caption{Common members lifetime in CR-DRL method versus ECRDP method.}
    \label{Fig_5:CML}    
\end{figure}
The stability of cluster heads and members in CR-DRL is evaluated using three metrics cluster head lifetime, cluster member lifetime, and cluster head change rate to ensure efficient routing. Stable clusters minimize disruptions, reduce re-clustering overhead, and enhance data transmission reliability, with longer cluster head lifetimes indicating balanced resource allocation and lower change rates reflecting steady clusters, reduced energy use, and improved packet delivery efficiency.
A recent paper dealing with stability of cluster heads and members is ECRDP \cite{A41}. As mentioned in the related work section, this method uses a machine learning approach with dynamic clustering and is a suitable candidate for comparison with the CR-DRL method. Figures \ref{Fig_5:CML}, \ref{Fig_6:CHL}, and \ref{Fig_7:CHCR} illustrate that CR-DRL surpasses ECRDP in common members lifetime, cluster head lifetime, and cluster head change rate, particularly in high-density scenarios where CR-DRL shows improvements of over 14\%, 7\%, and 21\%, respectively. The proposed CR-DRL approach achieves a higher cluster head lifetime and a lower cluster head change rate primarily due to the integration of the AC algorithm and the ADT mechanism. The AC algorithm enables the dynamic selection of cluster heads based on critical metrics such as encounter history, residual energy, and buffer capacity. This targeted selection ensures that nodes with higher stability and optimal resource availability are chosen as cluster heads, leading to prolonged cluster head lifetimes. Furthermore, the ADT mechanism adjusts clustering thresholds according to network density, ensuring clusters are neither too large nor too sparse. In high-density scenarios, smaller clusters reduce overhead, while in low-density conditions, larger clusters maintain connectivity. This dynamic adaptability minimizes the frequency of re-clustering events, thereby reducing the cluster head change rate.
Also, in this scenario, we conducted a comprehensive performance evaluation of the proposed CR-DRL protocol in comparison with the benchmark EEHCHR protocol \cite{New1}. To ensure a rigorous and equitable comparison, the evaluation was conducted using several key performance indicators: A round corresponds to a complete simulation cycle in which clustering, data transmission, and energy updates are executed across all nodes. First Node Dies (FND) denotes the round at which the first node depletes its energy, serving as an indicator of the network’s stability period. Half of the Nodes Die (HND) represents the round when 50\% of the deployed nodes have exhausted their energy, thus reflecting the broader network lifetime. The metric Alive Nodes quantifies the percentage of nodes that remain functional as the simulation progresses, while Residual Energy indicates the proportion of remaining total network energy relative to the initial energy budget. 
\begin{figure}[!t]
    \centering
    \includegraphics[width=2.5in]{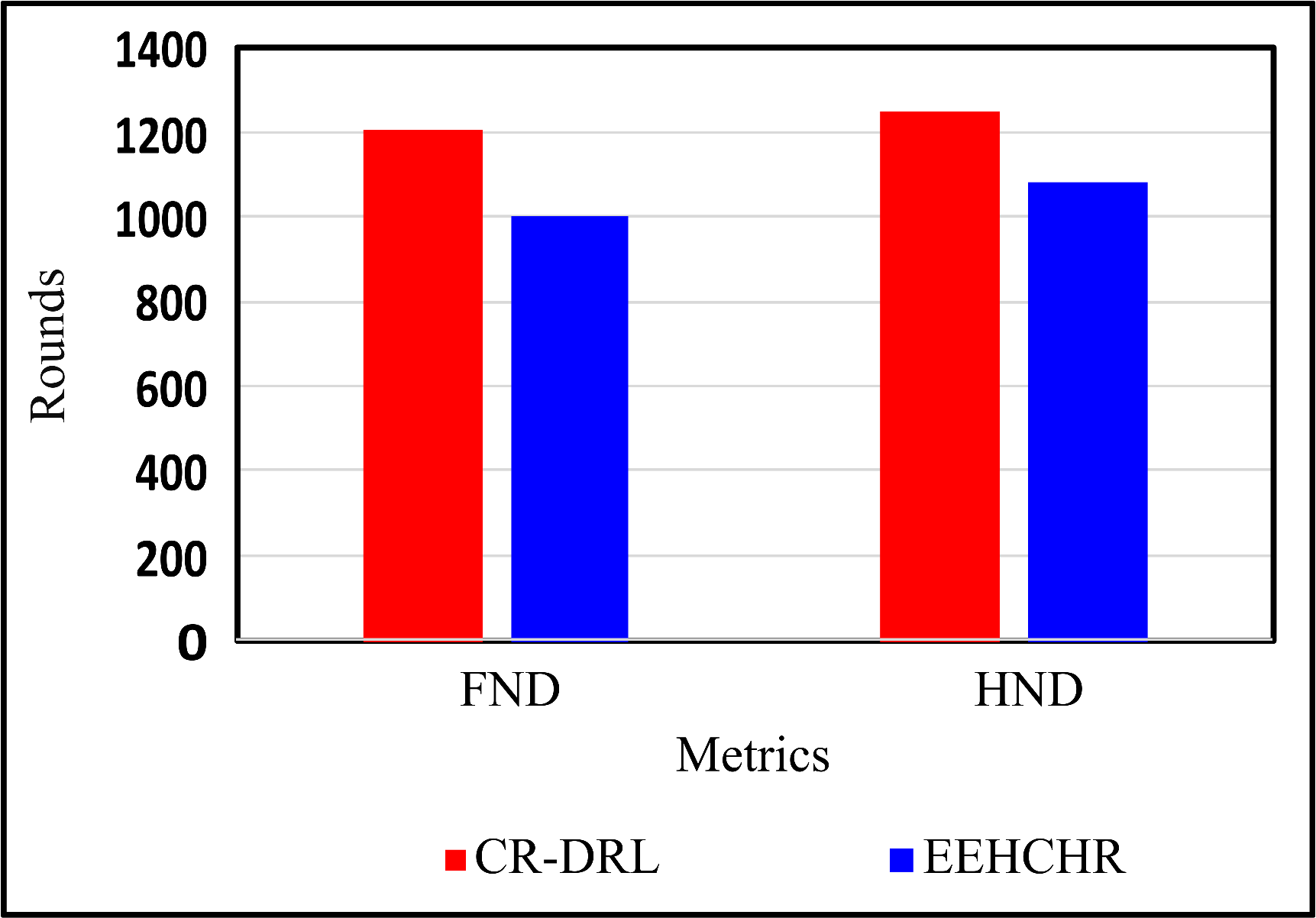}
    \caption{FND and HND comparison between CR-DRL and EEHCHR.}
    \label{FigNew}
\end{figure}
Figure \ref{FigNew} demonstrates that CR-DRL significantly delays the FND and HND compared with EEHCHR, achieving lifetime extensions of up to 20\% and 15.38\%, respectively. These gains stem from the integration of residual energy into the cluster head selection process, which prevents fragile nodes from being overused, and from the AC framework that balances forwarding tasks by considering energy, encounter history, and buffer capacity. As a result, critical nodes remain operational for longer, delaying the onset of large-scale node depletion that undermines network performance. Figure \ref{FigNew1} further demonstrates that CR-DRL maintains complete node operational capacity up to round 1400, whereas EEHCHR begins to experience node failures from round 1000 onward. By the conclusion of the simulation, CR-DRL retains up to 15\% of the nodes, in contrast to EEHCHR, which approaches near total depletion. To ensure a fair comparison of energy consumption, percentage-based normalization was applied to the residual energy measurements of both protocols.
\begin{figure}[!t]
    \centering
    \includegraphics[width=2.5 in]{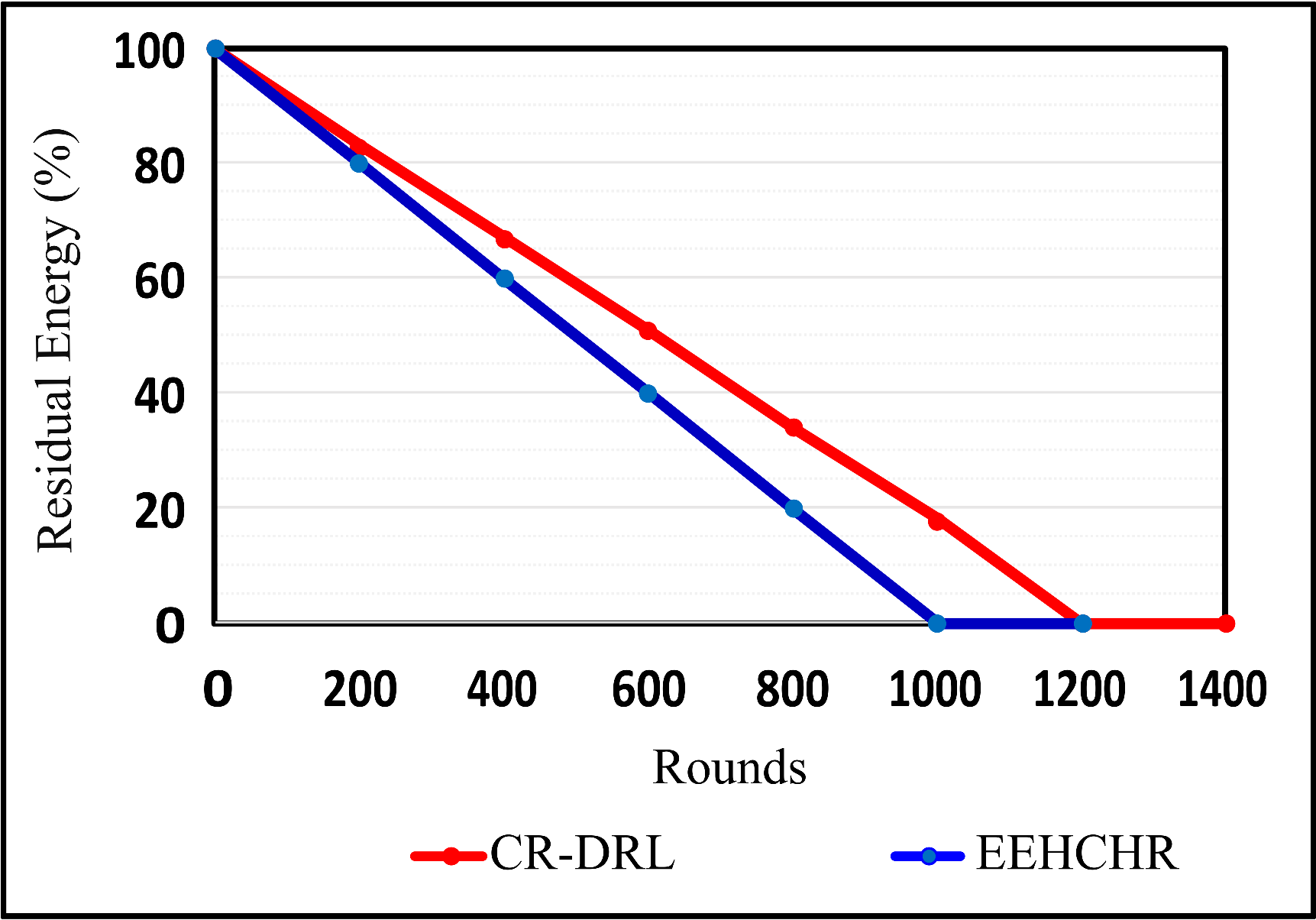}
    \caption{Residual energy comparison between CR-DRL and EEHCHR.}
    \label{FigNew2}
\end{figure}
As shown in Figure \ref{FigNew2}, CR-DRL demonstrates superior energy efficiency, retaining up to 17\% of the network’s residual energy at round 1000, while EEHCHR is fully depleted. This advantage results from CR-DRL’s residual energy–based cluster head selection, which allocates forwarding tasks to nodes with higher remaining energy, together with enhanced cluster stability that minimizes control overhead and avoids unnecessary energy dissipation.
\subsection{Second scenario}
In the second scenario, as noted above in Section \ref{Sec_Intro}, the CR-DRL approach allows TCP/IP routing protocols to be compatible with OppNets. Through adaptive clustering, the CR-DRL method can provide a stable connection between nodes that this connection is suitable for the TCP/IP protocol. We compare CR-DRL with SCF-Architecture and then compare the CR-DRL method with existing machine learning routing protocols defined in sources such as Mobility Adaptive \cite{S1}, RLProph \cite{S2}, and Latency‐Aware \cite{S3}. The evaluation parameters are average Hop count, average Delivery ratio, Throughput, E2E delay and average remaining energy.
\subsection{Experimental results} \label{Result}
\begin{figure}[!t]
    \centering
    \includegraphics[width=2.5in]{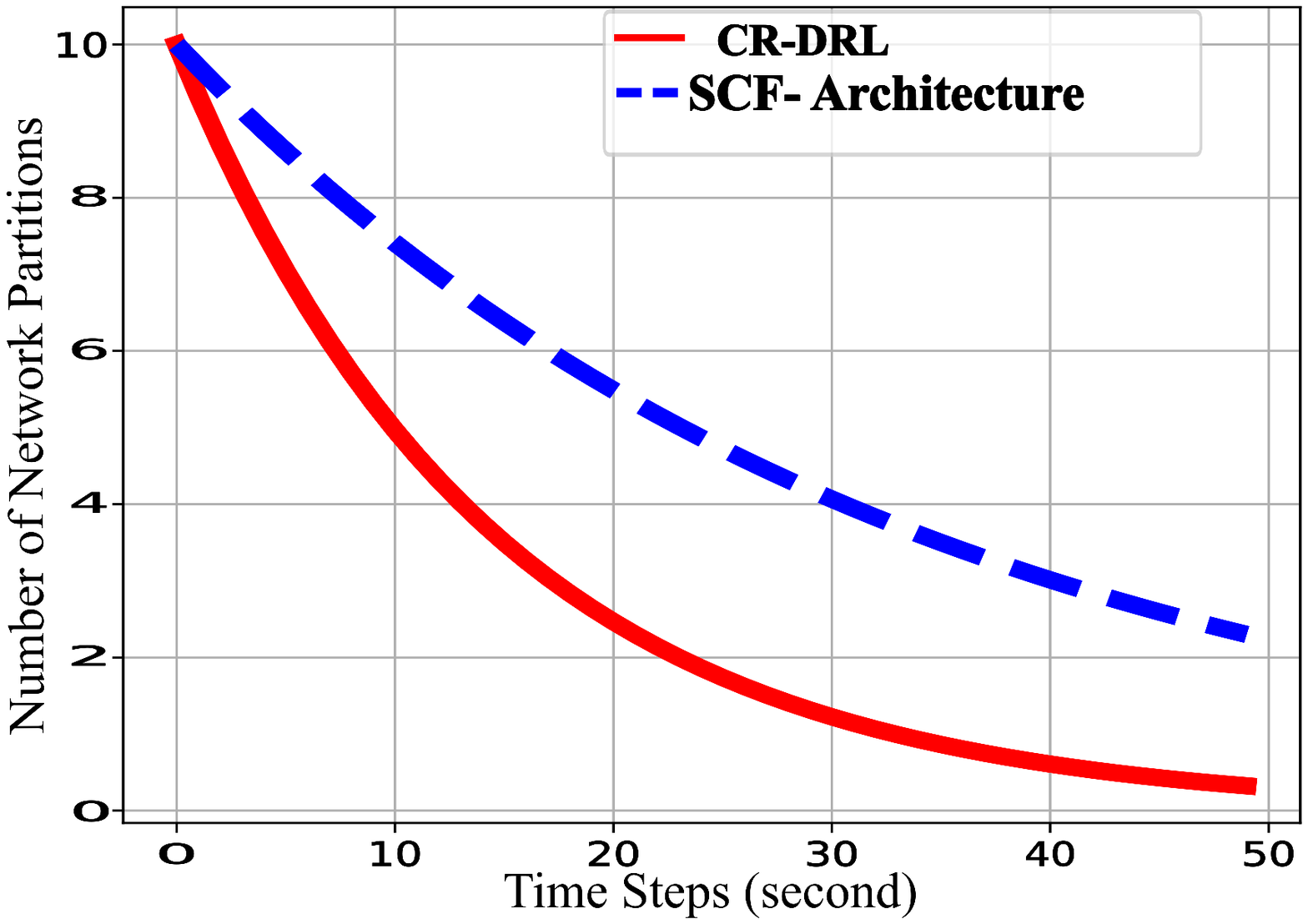}
    \caption{Partitioning of CR-DRL networks vs. SCF-Architecture.}
    \label{Fig_8:NP}
\end{figure}

\begin{figure}[!t]
    \centering
\includegraphics[width=2.5in]{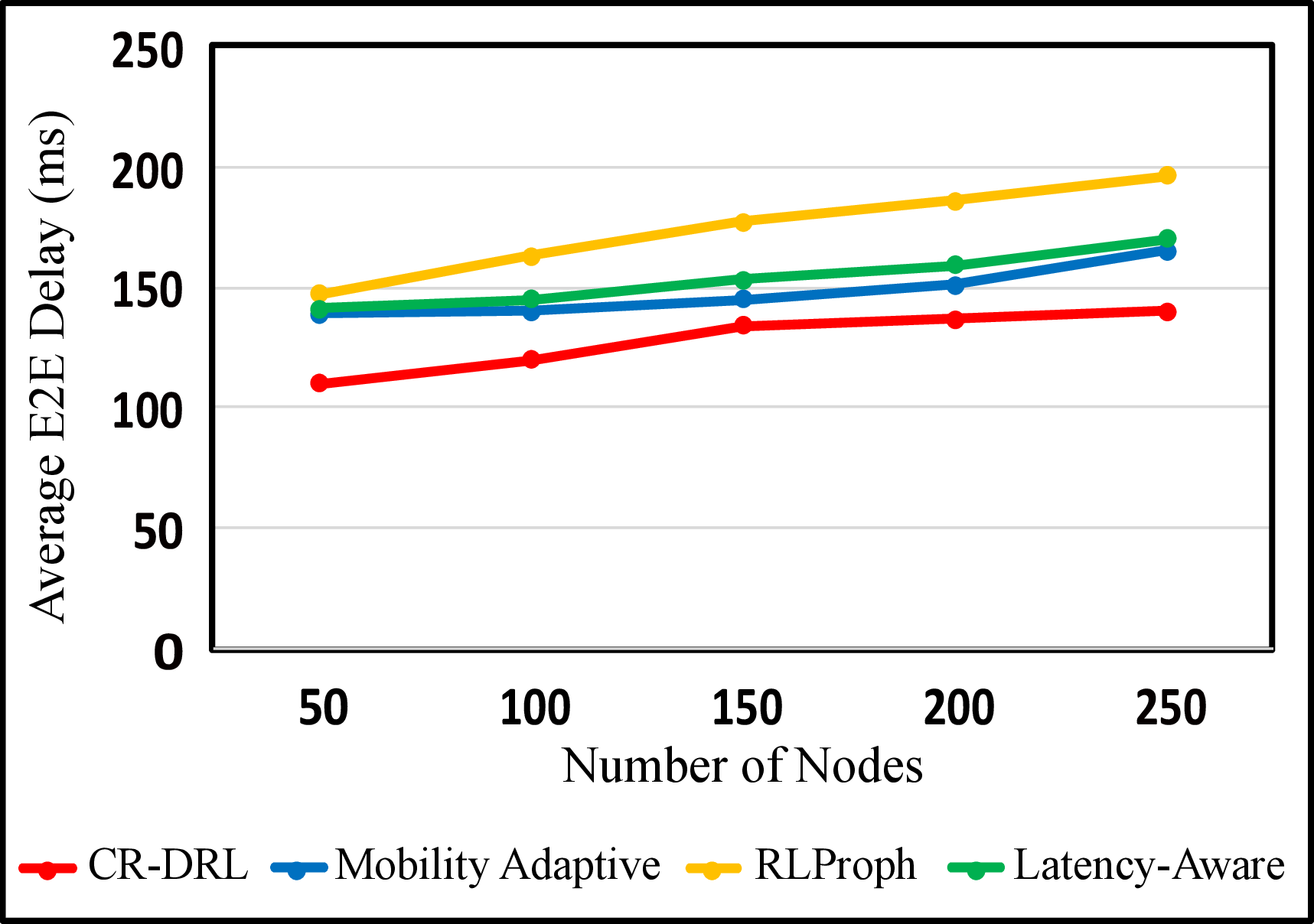}
    \caption{Average E2E delay across different node densities.}
    \label{Fig_9:EtoE}
\end{figure}

\begin{figure}[!t]
    \centering
    \includegraphics[width=2.5in]{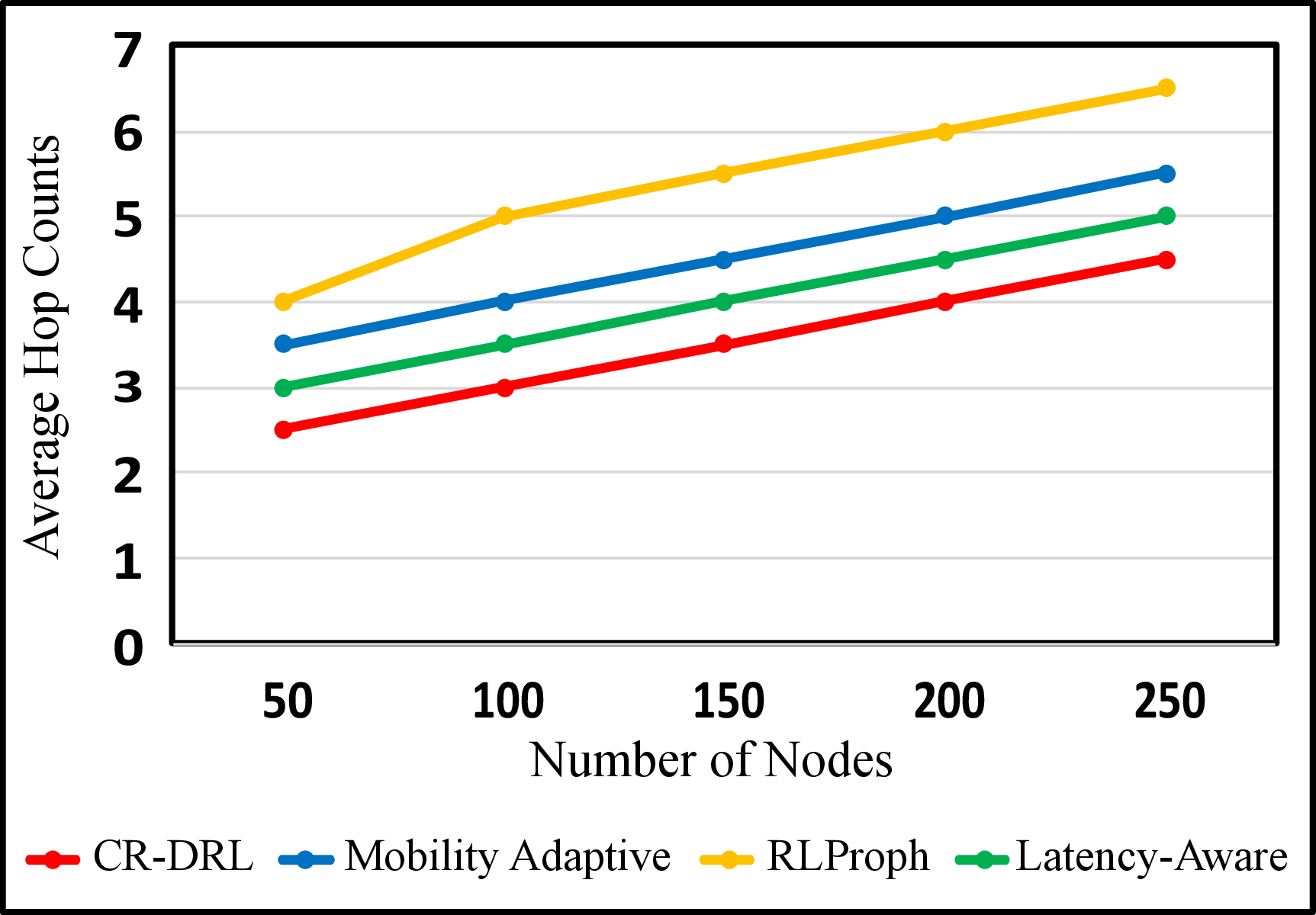}
    \caption{Average hop counts across different node densities.}
    \label{Fig_10:HC}
\end{figure}

\begin{figure}[!t]
    \centering\includegraphics[width=2.5in]{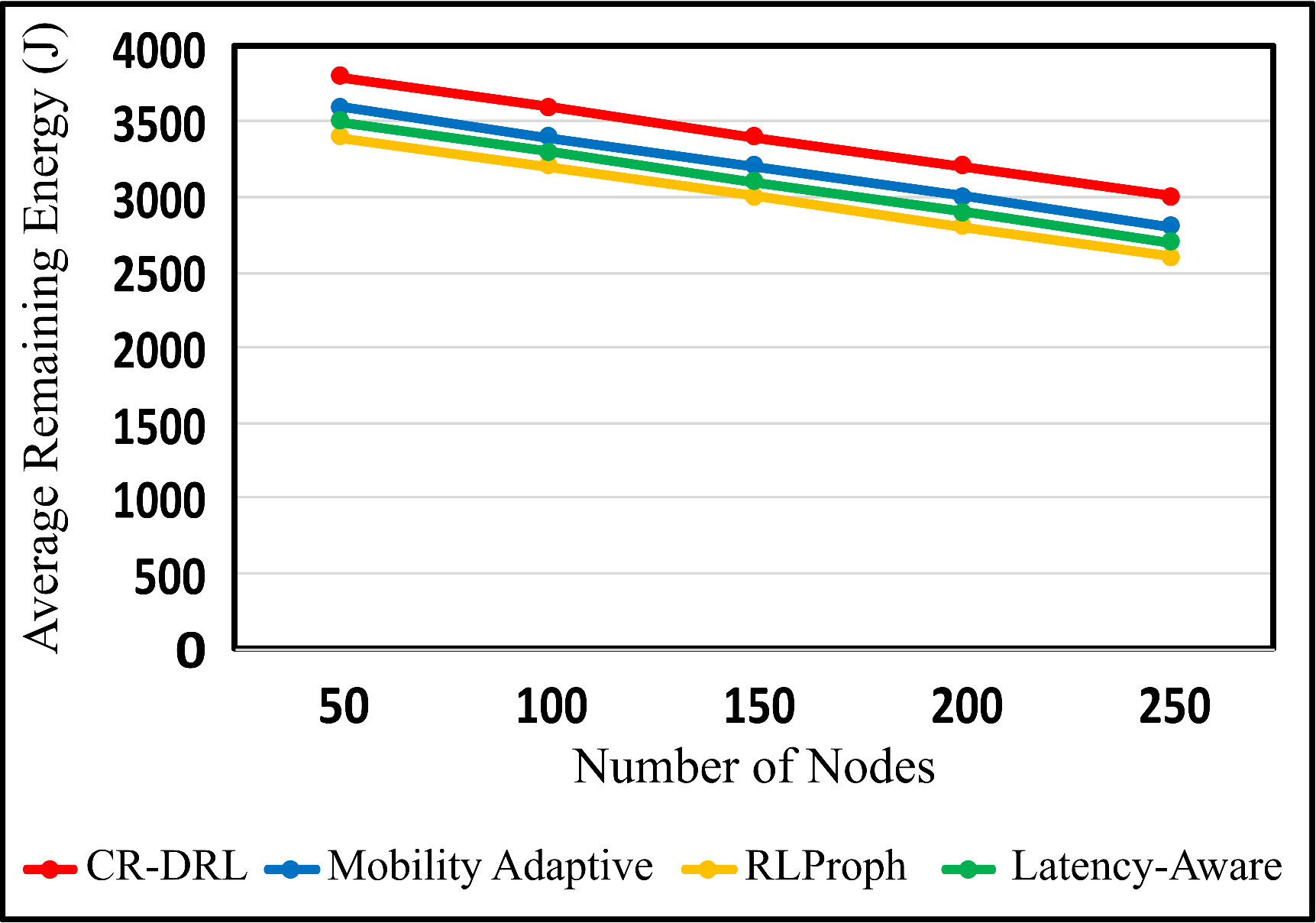}
    \caption{Average remaining energy in network.}
    \label{Fig_11:RE}
\end{figure}

Figure \ref{Fig_8:NP} illustrates the number of partitions in the network over time, demonstrating the improved performance of CR-DRL regarding network stability. Both CR-DRL method and SCF-Architecture have high partitioning at the initial point, almost with 9 and 10 partitions, respectively. However, over time, CR-DRL significantly reduces partitioning and achieves a connected network with only 1 partition at 35 seconds. On the other hand, SCF-Architecture continue to experience multiple partitions (around 3-6 partitions even after 50 seconds) and hence are causing the data transfer process get seriously hindered. CR-DRL reduces partitions in the network by 85\% compared to SCF-Architecture. Hence, the CR-DRL can maintain long-term connections and prevent network breakdowns, which can be suitable for TCP/IP networks.
\begin{figure}[!t]
    \centering
    \includegraphics[width=2.5in]{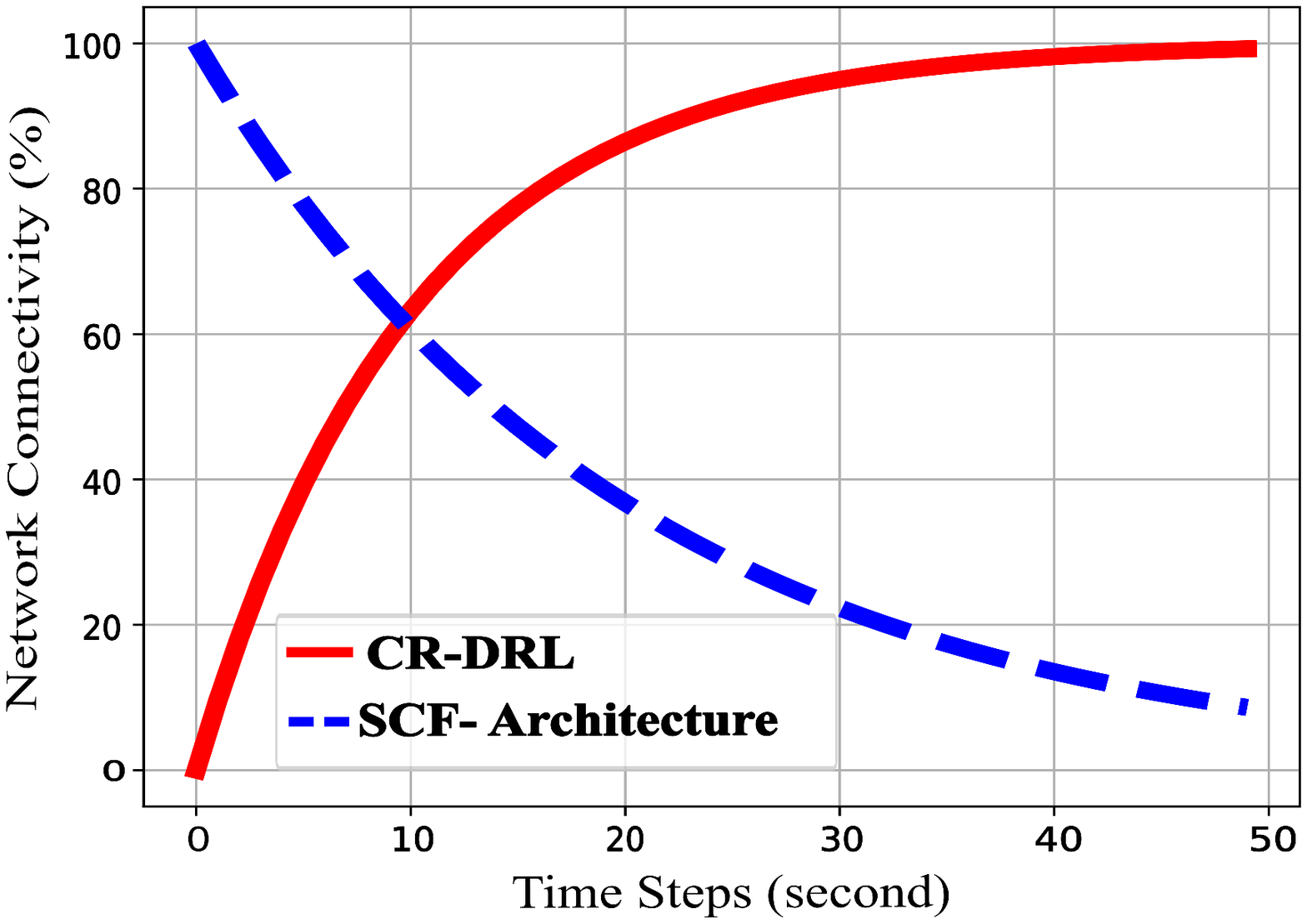}
    \caption{The connectivity rate of CR-DRL compared to SCF-Architecture.}
    \label{Fig_10:NC}
\end{figure}
\begin{figure}[!t]
    \centering
\includegraphics[width=2.5in]{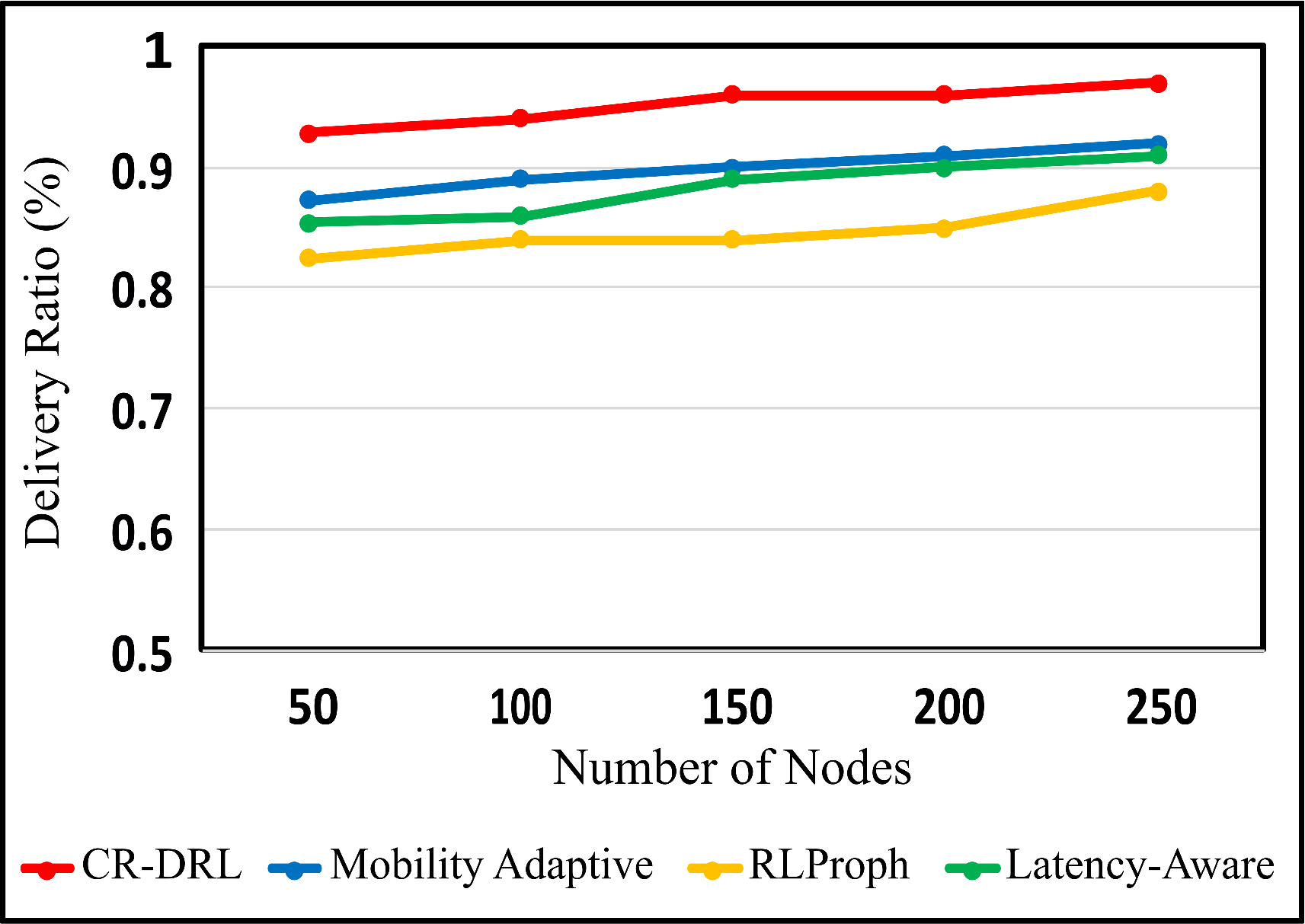}
    \caption{Delivery ratio across different node densities.}
    \label{Fig_8:DR}
\end{figure}
Figure \ref{Fig_10:NC} shows the percentage of network connectivity over time between CR-DRL and SCF-Architecture. SCF-Architecture initially show a higher percentage of network connectivity than CR-DRL. This is due to the inherent nature of SCF routing, in which nodes exchange packets at random encounters, causing a short-term spike in connectivity. This connectivity is transient as SCF-based methods do not maintain long-term connections nodes frequently disconnect, causing constant interruptions. On the other hand, CR-DRL starts with a slower connectivity rate as it requires an initial learning phase to establish stable clusters. CR-DRL is characterized by a rapid increase in connectivity at 80\% in the initial 20 seconds, whereas SCF-Architecture only achieve 40\% during the same time. Upon simulation termination (50 seconds), CR-DRL reaches near complete connectivity, while SCF-based networks are still fragmented at 58\%. This result demonstrates how CR-DRL efficiently suppresses network fragmentation via stable, dynamic cluster formation that offers an always-on network connectivity between nodes, which is critical in the TCP/IP-based communications.\\
Figure \ref{Fig_8:DR} demonstrates that the proposed CR-DRL protocol consistently achieves the highest packet delivery ratio across all network densities when compared with Mobility Adaptive, RLProph, and Latency-Aware Routing. The improvement is particularly pronounced in dense networks (250), where CR-DRL delivers more than 5\% and 10\% higher packets than Mobility Adaptive and RLProph, respectively. Additionally, CR-DRL was better than Latency‐Aware, with improvements of more than 6\%. This outcome can be attributed to two complementary factors: (i) cluster head selection that explicitly avoids nodes with limited energy or buffer capacity, and (ii) the ADT heuristic, which preserves connectivity by dynamically regulating cluster sizes. In contrast, baseline methods optimize only a single aspect (e.g., delay in Latency-Aware or density in Mobility-Adaptive) and are unable to sustain robustness across diverse operating conditions.
Figure \ref{Fig_9:EtoE} illustrates the average E2E delay across different densities. The CR-DRL approach consistently outperforms all baseline protocols, yielding notable reductions in packet delay up to 25\% in sparse topologies and reaching 28.5\% under dense network conditions. Stable clusters limit route disruptions, whereas selecting cluster heads by buffer capacity prevents congestion at overloaded nodes. In comparison, Latency-Aware routing reduces delay in low densities however, it suffers under high traffic because it ignores energy and buffer dynamics that lead to bottlenecks. RLProph and Mobility-Adaptive incur further delay from unstable forwarding paths and excessive re-routing. Importantly, CR-DRL’s reduction in delay does not come at the cost of delivery ratio or throughput, demonstrating a balanced optimization.
As shown in Figure \ref{Fig_10:HC}, CR-DRL achieves the lowest average hop count across all densities. At 250 nodes, CR-DRL reduces the number of hops by more than 18\% relative to Mobility Adaptive, up to 30\% compared to RLProph, and up to 10\% compared to Latency-Aware routing. The efficiency arises from packets being transmitted through shared member links between clusters, rather than relying on extended chains of opportunistic relays. In contrast, RLProph and Mobility-Adaptive introduce higher hop counts due to probabilistic and mobility-driven forwarding that fail to prioritize paths. Although Latency-Aware decreases hop counts, it lacks structural mechanisms, e.g., ADT, to avoid redundant inter-cluster transitions.
\noindent 
A reduction in hop count signifies shorter transmission distances between nodes, resulting in decreased energy usage for relaying messages within the network. 
The results presented in Figure \ref{Fig_11:RE} show that CR-DRL surpasses other Mobility Adaptive, RLProph, and Latency-Aware methods in terms of energy efficiency. Specifically, in high-density nodes (250), the CR-DRL approach enhances remaining energy by more than 7\%, 15\%, and 11\% compared to the mentioned methods. This performance results from three key mechanisms. First, intelligent cluster head selection considers residual energy levels to prevent depleted nodes from forwarding tasks. Second, the AC method distributes workload optimally to avoid node overburdening. Third, the ADT mechanism minimizes energy consumption by reducing frequent clustering processes. In contrast, baseline methods lack explicit energy considerations. Mobility Adaptive and RLProph employ probabilistic forwarding without considering node energy states, while Latency-Aware prioritizes delay reduction at the expense of energy consumption, resulting in accelerated network-wide battery depletion.
\begin{figure}[!t]
    \centering
\includegraphics[width=2.5in]{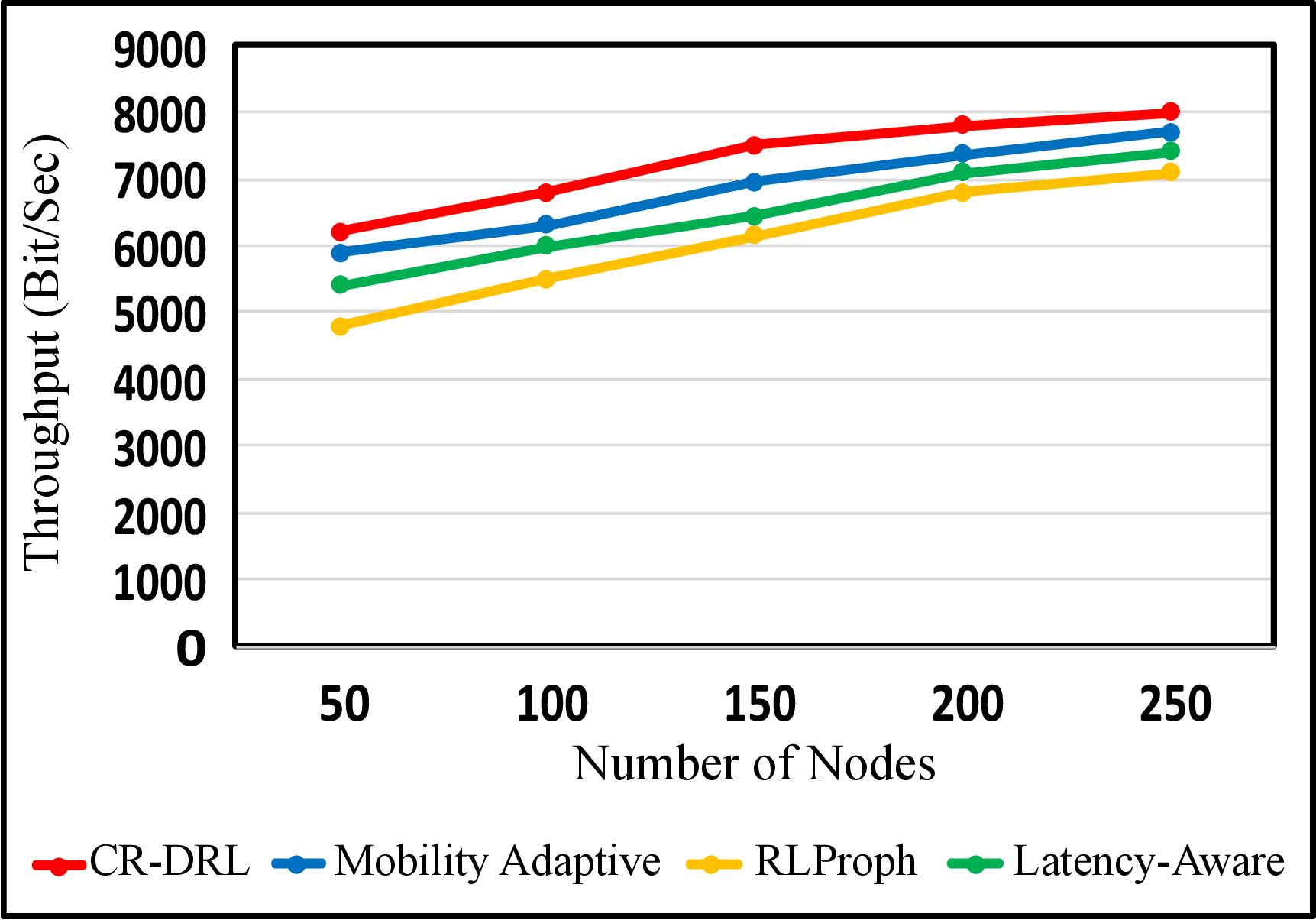}
    \caption{Average throughput across different node densities.}
    \label{Fig_12:Throughput}
\end{figure}
Throughput performance, depicted in Figure \ref{Fig_12:Throughput}, confirms the cumulative advantage of CR-DRL. At 250 nodes, throughput improves by 3\% over Mobility Adaptive, 12\% over RLProph, and 7\% over Latency-Aware routing. In contrast, RLProph is hindered by computational overhead that degrades performance, Mobility Adaptive protocol incurs packet losses arising from unstable routing paths under high vehicular mobility and rapidly varying topologies \cite{S1}, while Latency-Aware approach suffers congestion as traffic intensity increases.
\section{Conclusion and future work} \label{conclusion and future}

The paper proposes CR-DRL, a cluster-based routing protocol that leverages deep reinforcement learning to improve energy efficiency, delivery performance, and E2E delay in OppNets. By employing an AC framework for cluster head selection based on residual energy, buffer size, and encounter frequency, alongside an ADT function for dynamic cluster sizing, CR-DRL consistently outperforms benchmark methods. Simulation results demonstrate substantial gains over state-of-the-art approaches, with CR-DRL extending node lifetimes by up to 21\%, reducing overall energy consumption by 17\%, and keeping nodes active 15\% longer. Communication performance is also enhanced, achieving up to 10\% higher delivery ratio, 28.5\% lower latency, 7\% higher throughput, and a 30\% reduction in transmission steps required across the network. In future research, we will develop a framework that leverages generative AI to create synthetic mobility patterns and employs federated learning to optimize routing within 6G-enabled edge computing environments. This approach will enable real-time, efficient, and scalable intelligent transportation systems by integrating generative AI with next-generation network technologies.
\bibliography{CR-DRL}
\vspace{-0.2in}
\begin{IEEEbiography}
[{\includegraphics[width=1in,height=1.25in,clip,keepaspectratio]{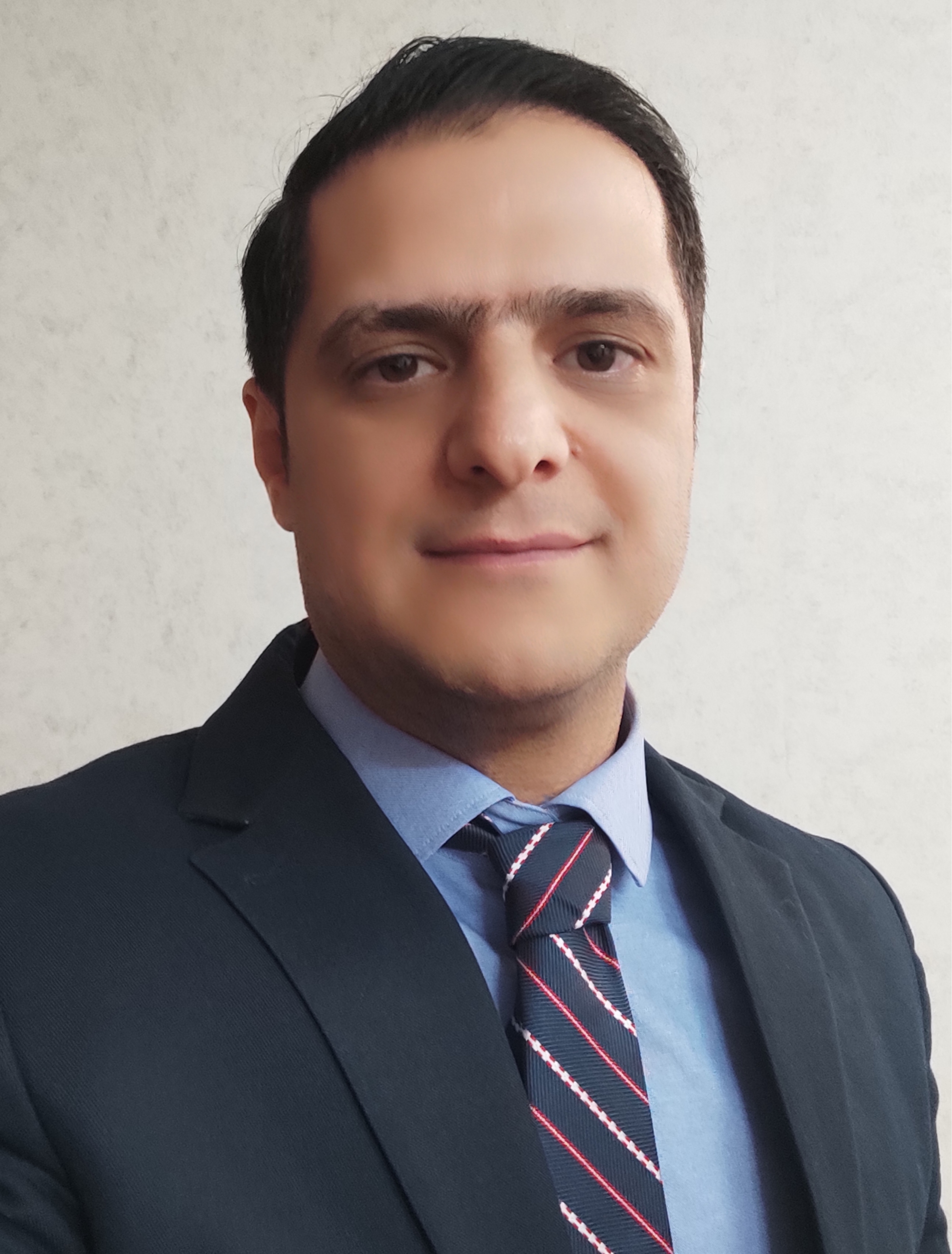}}]{Meisam Sharifi Sani} received his M.Sc. degree in Software Engineering from Iran in 2014. He is currently pursuing his Ph.D. degree in Electrical and Telecommunication Engineering at the University of Wollongong, Australia, under the Research Training Program (RTP) scholarship. His fields of research include Intrusion Detection Systems (IDS), Internet of Things (IoT), Internet of Vehicle (loV), Machine Learning, VDTN, and Wireless Networks.\end{IEEEbiography}
\vspace{-0.2in}
\begin{IEEEbiography}[{\includegraphics[width=1in,height=1.25in,clip,keepaspectratio]{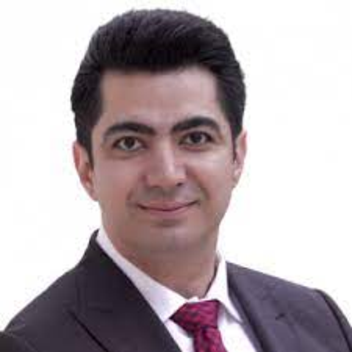}}]{Saeid Iranmanesh}
received the B.E. and M.S. degrees in computer science from Iran, and the Ph.D. degree in electrical and telecommunication engineering from the University of Wollongong, Wollongong, NSW, Australia, in 2015. He is currently a Senior Research Fellow with the School of Electrical, Computer and Telecommunication Engineering, University of Wollongong. He is a senior member of IEEE and Fellow of Engineers Australia. His research interests include wireless networks, vehicular networks, intelligent transportation systems, and smart cities.\end{IEEEbiography}
\vspace{-0.2in}
\begin{IEEEbiography}[{\includegraphics[width=1in,height=1.25in,clip,keepaspectratio]{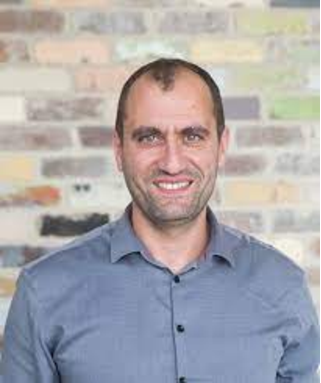}}]{Raad Raad}received a Bachelor of Engineering degree (Hons.1) in Electrical Engineering from the University of Wollongong, Australia in 1997, and a Ph.D. degree entitled Neuro-Fuzzy Logic Admission Control in Cellular Mobile Networks, in 2006. Since 2004, he has been with the School of Electrical, Computer and Telecommunications Engineering, University of Wollongong and he is the current Head of School. His current research interests include wireless communications, CubeSat, the IoT, and Antenna design in addition to advanced power systems. He is part of the ARC ITTC for Future Grids and ARC hub for Connected Health Sensors, both major research initiatives.
\end{IEEEbiography}
\vspace{-0.2in}
\begin{IEEEbiography}[{\includegraphics[width=1in,height=1.25in,clip,keepaspectratio]{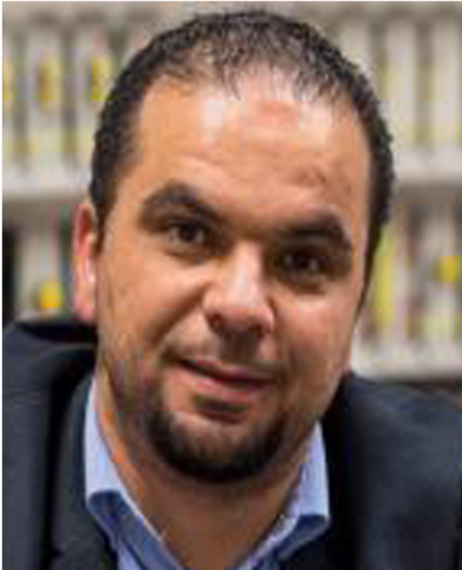}}]{Faisel Tubbal} received his B.E. in Electronic Technology from Tripoli, Libya (2004), and M.S. degrees in Telecommunication Engineering (2012) and Engineering Management (2013), as well as a Ph.D. in Telecommunication Engineering (2017) from the University of Wollongong, Australia. He began his career at the Libyan Centre for Remote Sensing and Space Science and later joined the University of Wollongong, where he is currently Laboratory Manager and Work Integrated Learning Coordinator. Since 2017, he has also served as Unit Convenor at Western Sydney University. He has authored a book, three chapters, and over 65 publications. His research focuses on CubeSat, wearable, and medical antennas, metamaterials, metasurfaces, and wireless power transfer. He is a recipient of multiple teaching excellence awards, including OCTAL, and a WATTLE fellow.

\end{IEEEbiography}
\vfill
\end{document}